\documentclass[twocolumn,english,prl]{revtex4-1}
\usepackage[T1]{fontenc}
\usepackage[latin9]{inputenc}
\usepackage{amsmath}
\usepackage{amssymb}

\makeatletter
\usepackage{braket}

\makeatother

\usepackage{babel}
\begin{document}
\title{Quantum Dynamics with Electronic Friction}
\author{Rocco Martinazzo$^{1,2,*}$, Irene Burghardt$^{3}$}
\affiliation{$^{1}$Department of Chemistry, Università degli Studi di Milano,
Via Golgi 19, 20133 Milano, Italy}
\email{rocco.martinazzo@unimi.it}

\affiliation{$^{2}$Istituto di Scienze e Tecnologie Molecolari, CNR, via Golgi
19, 20133 Milano, Italy}
\affiliation{$^{3}$Institute of Physical and Theoretical Chemistry, Goethe University
Frankfurt, Max-von-Laue-Str. 7, D-60438 Frankfurt/Main, Germany}
\begin{abstract}
A theory of electronic friction is developed using the exact factorization
of the electron-nuclear wavefunction. No assumption is made regarding
the electronic bath, which can be made of independent or interacting
electrons, and the nuclei are treated quantally. The ensuing equation
of motion for the nuclear wavefunction is a non-linear Schr\"{o}dinger
equation including a friction term. The resulting friction kernel
agrees with a previously derived mixed quantum-classical result by
Dou, Miao \& Subotnik (\emph{Phys. Rev. Lett.} \textbf{119}, 046001
(2017)), except for a \emph{pseudo}-magnetic contribution in the latter
that is here removed. More specifically, it is shown that the electron
dynamics generally washes out the\emph{ gauge} fields appearing in
the adiabatic dynamics. However, at T=0 K, the \emph{pseudo}-magnetic
force is fully re-established in the typical situation where the electrons
respond rapidy on the slow time-scale of the nuclear dynamics (Markov
limit). Hence, we predict Berry's phase effects to be observable also
in the presence of electronic friction, and non-trivial geometric
phases should be attainable for molecules on metallic magnetic surfaces. 
\end{abstract}
\maketitle
\textbf{\emph{Introduction}}. When molecules scatter off, react or
simply vibrate at metal surfaces they can induce electronic excitations
in the substrate, in addition to the usual phonon excitations. This
energy loss mechanism is a clear violation of the Born-Oppenheimer
(BO) approximation and can give rise to intriguing phenomena, including
electron transfer processes and generation of chemically-induced currents
\cite{Nienhaus2002}. Often, though, electron excitation is of limited
extent and reduces to a frictional force of electronic origin that
acts on the molecular degrees of freedom, in addition to the usual
Born-Oppenheimer forces, a situation where the so-called ``BO dynamics
with electronic friction'' description is appropriate.

Electronic friction has a long history \cite{Dou2018}. One early
derivation of the frictional forces that electrons exert on a set
of (classically) moving nuclei is due to Head-Gordon and Tully \cite{Head-Gordon1995},
who first derived an expression for the relevant friction kernel $\gamma$.
The result of Ref. \cite{Head-Gordon1995}, 
\begin{equation}
\gamma_{kj}^{\text{HGT}}=\pi\hbar\sum_{ab}\braket{a|\partial_{k}h|b}\braket{b|\partial_{j}h|a}\delta(\epsilon_{a}-\epsilon_{F})\delta(\epsilon_{b}-\epsilon_{F})\label{eq:Head-Gordon Tully}
\end{equation}
where $a,b$ label single-particle states, $k$,$j$ label nuclear
degrees of freedom, $h$ is the one-particle Hamiltonian and $\epsilon_{F}$
the Fermi energy, was obtained at zero temperature in the independent
electron approximation, and found to be consistent with earlier results
on vibrational relaxation at metal surfaces \cite{Persson1980,Persson1982,Head-Gordon1992}.
It was later re-derived using different methodologies, including influence
functionals \cite{Brandbyge1995} and nonequilibrium Green\textquoteright s
functions \cite{Lu2012} (see Ref. \cite{Dou2018} for a comprehensive
account). This form of electronic friction, combined with \emph{first
principles} electronic structure theory \cite{Maurer2016,Jin2019,Zhang2019},
has been applied to a variety of problems \cite{Luntz2005,Luntz2006,Fuchsel2011,Monturet2010,Askerka2016,Zhang2019,Box2020},
and Langevin dynamics with electronic friction and density-functional
theory potentials is nowadays a standard tool to investigate the dynamics
of molecules at metal surfaces. Other works addressed the issue of
non-thermal, yet steady-state, electronic baths (\emph{e.g.,} current-carrying
metals) \cite{Bode2011,Bode2012} and of the electron-electron interactions
\cite{Daligault2007,Dou2017}, and showed the importance of going
beyond a mean-field treatment of the electronic dynamics \cite{Dou2017}.
In particular, Dou, Miao \& Subotnik (DMS), using a mixed quantum-classical
approach, derived a completely general friction kernel that applies
to interacting electrons, is valid out of equilibrium, and reduces
to previously published expressions for independent electrons \cite{Dou2017}.
DMS wrote the electronic friction tensor, in the Markov limit, as
\begin{equation}
\gamma_{kj}^{\text{DMS}}=-\int_{0}^{\infty}\text{tr}_{e}\left(\left(\partial_{k}H_{\text{el}}\right)e^{-\frac{i}{\hbar}H{}_{\text{el}}\tau}\left(\partial_{j}\rho\right)e^{+\frac{i}{\hbar}H{}_{\text{el}}\tau}\right)d\tau\label{eq:DMS friction}
\end{equation}
where $\rho$ is the steady-state electron density-operator, $H_{\text{el}}$
is the electronic Hamiltonian, possibly including interactions between
electrons, and $\text{tr}_{e}$ denotes the trace over the electron
degrees of freedom.

The electronic friction limit can be considered a first order ``realization''
of the adiabatic approximation in a situation where a continuum of
electronic states and the ensuing fast relaxation guarantee that,
on the time scale of nuclear motion, the electrons follows \emph{adiabatically}
the nuclei. However, the adiabatic approximation, in addition to the
usual BO forces, introduces \emph{gauge} fields that reflect the geometric
properties of the electronic eigenspaces when viewed as functions
of the slow parameters (the nuclear degrees of freedom), and it is
not clear whether and how these disappear when the electron dynamics
is taken into account. The answer to this question lies into the nuclear
wavefunction, since the electronic degrees of freedom are traced out
in the above electronic-friction description and the phase of the
electronic wavefunction cannot be tracked in any realistic experiment.
Hence, a full quantum description of the dynamics is made necessary,
and this is the purpose of the present Letter. 

The Letter is organized as follows. After summarizing some basic properties
of the adiabatic approximation we analyze the $T=0$ K exact quantum
dynamics of the combined electron-nuclear system using a representation
that closely resembles the adiabatic one. We shall show that the introduction
of the electron dynamics generally washes out the above mentioned
\emph{gauge} fields, and thus removes any Berry's phase effect from
the dynamics. Later, we analyze the case of an electronic bath that
relaxes quickly on the time-scale relevant for the nuclaer motion,
and derive an electronic-friction kernel that describes the corresponding
electronic-friction dynamical regime. In the Markov limit of a memoryless
friction, we shall show that the \emph{pseudo}-magnetic forces are
fully restored, thereby making geometric phase effects potentially
observable. Importantly, it is further shown how the adiabatic Hamiltonian
has to be modified to include friction in the nuclear dynamics, and
how the equation of motion for the nuclear wavefunction is turned
into a non-linear equation of the Schr\"{o}dinger-Langevin type.

\textbf{\emph{Adiabatic approximation.}} The adiabatic approximation
\cite{Born1927} and the related adiabatic theorem \cite{Born1928,Kato1950a}
has under-pinned research into quantum systems with slowly evolving
parameters, and form the basis of the theory of energy level crossings
in molecules, of the Gell-Mann\textendash Low theorem in quantum field
theory and of Berry\textquoteright s geometrical phase. When the slow
parameters $\mathbf{x}$ (here, the nuclear coordinates) are considered
as dynamical variables the approximation can be recast as a variational
approximation with the wavefunction \emph{ansatz}
\[
\ket{\Psi_{t}}=\int_{X}d\mathbf{x}\psi_{t}(\mathbf{x})\ket{u_{n}(\mathbf{x})}\ket{\mathbf{x}}
\]
where $\ket{u_{n}(\mathbf{x})}$ is a chosen time-independent ``frame''
of the $n^{\text{th}}$ electronic eigenstace, $\ket{\mathbf{x}}$
are position eigenstates of the slow variables, and $\psi(\mathbf{x})$
is the nuclear wavefunction. The latter satisfies the variational
equation of motion $H_{n}^{\text{eff}}\psi=i\hbar\partial_{t}\psi$
with the effective Hamiltonian
\begin{equation}
H_{n}^{\text{eff}}=\frac{1}{2}\sum_{ij}\xi^{ij}\hat{\pi}_{i}\hat{\pi}_{j}+\left(E_{n}-\hbar A_{0}+\phi\right)\label{eq:effective Hamiltonian}
\end{equation}
Here, $E_{n}=\braket{u_{n}|H_{\text{el}}|u_{n}}$ is the $n^{\text{th}}$
potential energy surface, $H_{\text{el}}\equiv H_{\text{el}}(\mathbf{x})$
the electronic Hamiltonian, $\xi^{ij}$ is a coordinate-independent
inverse-mass tensor of the slow variables and the mechanical momentum
for the $j^{\text{th}}$ degree of freedom $\hat{\pi}_{j}=\hat{p}_{j}-\hbar A_{j}$
is defined in terms of the Berry's connection\footnote{Henceforth, for notational convenience, we shall omit the $n$ dependence
of $A_{j}$, $\hat{\pi}_{j}$, etc. and of the quantum geometric tensor
$q_{ij}$ introduced below. } $A_{j}=i\braket{u_{n}|\partial_{j}u_{n}}$ and the coordinate representation
of the momentum operator $\hat{p}_{j}$. 

The geometric properties of the adiabatic approximation are subsumed
in the quantum geometric tensor \cite{Provost1980,Berry1984} 
\[
q_{ij}=\braket{\partial_{i}u_{n}|Q_{n}|\partial_{j}u_{n}}\ \ \ Q_{n}\equiv1-\ket{u_{n}(\mathbf{x})}\bra{u_{n}(\mathbf{x})}
\]
which determines both the scalar potential $\phi(\mathbf{x})$ and
the commutation properties of the $\hat{\pi}_{j}$'s, through its
real ($\Re q_{ij}=g_{ij}$) and imaginary ($\Im q_{ij}=-B_{ij}/2$)
parts, respectively, \emph{i.e.}, 
\[
\phi=\frac{\hbar^{2}}{2}\sum\xi^{ij}g_{ij}\ \ \ [\hat{\pi}_{i},\hat{\pi}_{j}]=i\hbar^{2}B_{ij}
\]
Here, $g_{ij}$ is a quantum metric and $B_{ij}$ is the $ij^{\text{th}}$
component of the Berry's curvature, \emph{i.e.}, the exterior derivative
of the linear differential form $\omega=\sum_{j}A_{j}dx^{j}$. Furthermore,
for later convenience, we have introduced a term $A_{0}=i\braket{u_{n}|\partial_{t}u_{n}}$
that allows more general, time-dependent \emph{gauge} transformations
($A_{0}\equiv0$ is the usual choice)\footnote{It is worth noticing that the transformation $\ket{u_{n}}\rightarrow e^{-iE_{n}t/\hbar}\ket{u_{n}}$
moves the BO PES into the kinetic energy term, since $\hbar A_{0}\equiv E_{n}$
in this \emph{gauge,} and $\hbar A_{k}\rightarrow\hbar A_{k}+(\partial_{k}E_{n})t$. }. 

The ensuing nuclear dynamics is governed by the Born-Oppenheimer force,
$F_{k}^{\text{BO}}=-\partial_{k}E_{n}$, in conjuction with a \emph{pseudo}-Lorentz
force comprising an electric 
\[
F_{k}^{\text{el}}=-\partial_{k}\phi=-\frac{\hbar^{2}}{2}\sum_{ij}\xi^{ij}\frac{\partial g_{ij}}{\partial x^{k}}
\]
and a magnetic contribution
\[
F_{k}^{\text{mag}}=\frac{\hbar}{2}\sum_{j}\left(\hat{v}^{j}B_{kj}+B_{kj}\hat{v}^{j}\right)
\]
($\hat{v}^{j}=\sum_{i}\xi^{ji}\hat{\pi}_{i}$ being the velocity operator)
that have a purely geometrical origin and represent the legacy of
the adiabatic constraint to the wavefunction. The above forces are
separately \emph{gauge} invariant. An additional \emph{gauge}-invariant
component related to the electron dynamics (ED) formally exists
\begin{equation}
F_{k}^{\text{ED}}=\hbar\left(\partial_{k}A_{0}-\partial_{t}A_{k}\right)=-2\hbar\Im\braket{\partial_{k}u|Q_{n}|\partial_{t}u}\label{eq:F_ED}
\end{equation}
but vanishes identically in the adiabatic approximation. 

As shown in the Supplemental Material (SM), the quantum geometric
tensor also determines the local-in-time error \cite{Martinazzo2020,Martinazzo2021}
of the adiabatic approximation. The latter takes approximately the
form of an expectation value of the quantized quantum geometric tensor
\[
\varepsilon^{2}\approx\bra{\psi}\sum_{ij}\hat{v}^{i}q_{ij}\hat{v}^{j}\ket{\psi}_{X}
\]
(where $X$ denotes integration w.r.t. the nuclear DOFs only) and
measures the tendency of the system to undergo a non-adiabatic transition
at short-time. In fact, $P_{\text{nad}}\approx\varepsilon^{2}(t-t_{s})^{2}$
is the total transition probability, if the adiabatic approximation
were suddently lifted at $t=t_{s}$. For comparison, we notice that
if the slow variables were simply some parameters, and not dynamical
variables, we would have exactly
\[
\varepsilon^{2}=\sum_{ij}V^{i}q_{ij}V^{j}\equiv\sum_{ij}V^{i}g_{ij}V^{j}
\]
where $V^{i}$ is the classical velocity of the $i^{\text{th}}$ parameter
(see SM for details).

\textbf{\emph{Exact dynamics}}. The \emph{exact} wavefunction can
be yet represented in a form similar to above,
\[
\ket{\Psi_{t}}=\int_{X}d\mathbf{x}\psi_{t}(\mathbf{x})\ket{u_{t}(\mathbf{x})}\ket{\mathbf{x}}
\]
but now with a time-dependent electronic state $\ket{u_{t}(\mathbf{x})}$
for each nuclear configuration $\mathbf{x}$. This is the so-called
exact-factorization of the wavefunction \cite{Abedi2010,Abedi2012},
which is an ``intermediate'' representation that can be obtained
by introducing a local basis of nuclear states $\{\ket{\mathbf{x}}\}$\emph{,}
and imposing a normalization condition on the ensuing local electronic
states,
\[
\braket{\mathbf{x}|\Psi_{t}}_{X}=\psi_{t}(\mathbf{x})\ket{u_{t}(\mathbf{x})}\ \braket{u_{t}(\mathbf{x})|u_{t}(\mathbf{x})}=1
\]
The corresponding equations of motion \cite{Abedi2010,Abedi2012}
are re-derived in SM using a projection-operator tenchique that emphasizes
their \emph{gauge} transformation properties. They can be summarized
as follows. The nuclear wavefunction satisfies $H^{\text{eff}}\psi=i\hbar\partial_{t}\psi$,
where the effective Hamiltonian takes precisely the form of Eq. \ref{eq:effective Hamiltonian},
but now $\ket{u(\mathbf{x},t)}$ replaces $\ket{u_{n}(\mathbf{x})}$
everywhere and $A_{0}=i\braket{u|\partial_{t}u}$ is an arbitrary
real \emph{guage} constraint that guarantees normalization of the
local electronic state. The ensuing equation of motion for the \emph{exact}
nuclear wavefunction are thus formally \emph{very} similar to the
adiabatic ones. The additional force, due solely to the electron dynamics,
is the \emph{gauge}-invariant term of Eq. \ref{eq:F_ED} that, differently
from the adiabatic approximation, is generally non zero. That is,
the total force reads \emph{exactly} as $F_{k}^{\text{tot}}=F_{k}^{\text{BO}}+F_{k}^{\text{el}}+F_{k}^{\text{mag}}+F_{k}^{\text{ED}}$,
where $F_{k}^{\text{BO}}\equiv-\partial_{k}\braket{u|H_{\text{el}}|u}=-\partial_{k}E_{\text{el}}$
is a time-dependent BO force, the geometric forces $F_{k}^{\text{el}}$
and $F_{k}^{\text{mag}}$ stem from the time-dependent geometric tensor
$q_{ij}$ and $F_{k}^{\text{ED}}$ describes the electron dynamics.

The electronic equation of motion takes the form
\begin{equation}
i\hbar Q\partial_{t}\ket{u}=QH_{\text{el}}\ket{u}+K[\psi_{t}]\ket{u}\label{eq:electronic equation}
\end{equation}
where the second term on the r.h.s. describes the `electron drag'
with the nuclear motion and reads as 
\begin{equation}
K[\psi_{t}]\ket{u(t)}=-i\hbar\sum_{j}V^{j}Q\ket{\partial_{j}u(t)}-\hbar R\ket{u(t)}\label{eq:drag}
\end{equation}
Here $V^{j}(\mathbf{x})=(\hat{v}^{j}\psi_{t})/\psi_{t}$ is the (\emph{gauge}-invariant)
time-dependent, complex-valued nuclear velocity field, $Q=1-\ket{u}\bra{u}$,
$R\ket{u}=\frac{\hbar}{2}\sum_{ij}\xi^{ij}D_{ij}\ket{u}$, and
\[
D_{ij}\ket{u}=iA_{i}Q\ket{\partial_{j}u}+iA_{j}Q\ket{\partial_{i}u}+Q\ket{\partial_{i}\partial_{j}u}
\]
is a second derivative of the electronic states. In this form the
effective electronic Hamiltonian is the sum of \emph{gauge} tensorial
terms $G$, \emph{i.e.}, of terms that behave simply as $G\ket{u}\rightarrow e^{-i\varphi}G\ket{u}$
under the \emph{gauge} transformation $\ket{u}\rightarrow e^{-i\varphi}\ket{u}$.
This Hamiltonian can be used to include explicitly the electron reaction
into the nuclear equation of motion. Specifically, plugging $Q\partial_{t}\ket{u}$
into $F_{k}^{\text{ED}}$ above one obtains a correction to the previous
\emph{pseudo}-electric and \emph{pseudo}-magnetic forces, that are
turned into
\begin{align}
F_{k}^{\text{el,c}} & =2\hbar\sum_{j}g_{kj}\Im V^{j}\nonumber \\
 & -\hbar^{2}\sum_{ij}\xi^{ij}\left(\Re\braket{\partial_{i}u|D_{kj}u}+\Re\braket{\partial_{k}u|D_{ij}u}\right)\label{eq:F_el corrected}
\end{align}
\begin{equation}
F_{k}^{\text{mag,c}}=\frac{\hbar}{2}\sum_{j}\left(\hat{v}^{j}B_{kj}+B_{kj}\hat{v}^{j}\right)-\hbar\sum_{j}B_{kj}\Re V^{j}\label{eq:F_mag corrected}
\end{equation}
\emph{plus} a genuine non-Born-Oppenheimer term 
\[
F_{k}^{\text{NBO}}=2\Re\braket{\partial_{k}u|QH'_{\text{el}}|u}
\]
where $H'_{\text{el}}=H_{\text{el}}-E_{\text{el}}$ for later convenience.
The latter force vanishes identically when $\ket{u}$ is an eigenstate
of $H_{\text{el}}$ and, more generally, is bound by the size of the
(local) energy fluctuations in the electronic subsystem, $\Delta E_{\text{el}}^{2}=\braket{\left(H_{\text{el}}-E_{\text{el}}\right)^{2}}$,
through the $k^{\text{th}}$ diagonal component of the quantum geometric
tensor, namely $|F_{k}^{\text{NBO}}|\leq2\Delta E_{\text{el}}\sqrt{g_{kk}}$.
Thus, the total force acting on the $k^{\text{th}}$ nuclear degree
of freedom can be written as
\[
F_{k}^{\text{tot}}=F_{k}^{\text{BO}}+F_{k}^{\text{el,c}}+F_{k}^{\text{mag,c}}+F_{k}^{\text{NBO}}
\]
and represents the same total force given above but now explicitly
including the electron reaction. The key point about this expression
is that, as shown in SM, the dynamically corrected \emph{pseudo}-Lorentz
force vanishes \emph{identically} when averaged over an arbitray nuclear
state since
\[
\braket{\psi|F_{k}^{\text{el,c}}|\psi}_{X}=\braket{\psi|F_{k}^{\text{mag,c}}|\psi}_{X}=0
\]
and thus it disappears in the quantum-classical limit where the electronic
friction approximation is often invoked. Hence, electronic friction
must come from the term $F_{k}^{\text{NBO}}$. 

We stress that the above results are \emph{exact} and tell us that,
in general, the effect of the electron dynamics is to wash out the
\emph{pseudo}-Lorentz force appearing in the adiabatic limit. As shown
in SM, the vanishing of the average \emph{pseudo}-electric force expresses
conservation of the quantum metric, since such force takes the form
of a(n expectation value of the) Ricci-Levi Civita covariant derivative
of the metric tensor. This appears reasonable since the quantum metric
and the related \emph{pseudo}-electric force measure the error in
the adiabatic \emph{approximation} while the dynamics considered here
is exact. Likewise, the vanishing of the \emph{pseudo-}magnetic force
signals the quenching of geometric phase effects that, in fact, should
disappear when the dynamics is far from the adiabatic regime\footnote{We remark that the adiabatic approximation does \emph{not} correspond
to $F_{k}^{\text{NBO}}=0$ \--- that would leave the electron corrections
to the forces \--- rather to $F_{k}^{\text{ED}}=0$. In a sense,
the Born-Oppenheimer approximation incorporates these corrections
and thus accounts partially for the electron reaction, thereby removing
geometric phase effects and the repulsive \emph{pseudo}-electric forces
that become important at conical intersections.}. 

\textbf{\emph{Electronic friction.}} Let us now focus on the electronic
equation Eq. \ref{eq:electronic equation} in the situation where
the electronic system relaxes quickly to the ground-state $\ket{u_{0}}$
and the deviation $\ket{\Delta u}=\ket{u(t)}-\ket{u_{0}(t)}$ remains
small throughout the nuclear dynamical evolution (here, $\ket{u_{0}(t)}$
is the time-evolving ground electronic state). This is the condition
where linear response theory (LRT) applies and also the situation
where the electronic-friction picture is appropriate (see SM for details).
Under these circumstances, $\braket{\Delta u|u_{0}}\equiv0$, the
BO forces $F_{k}^{\text{BO}}$ acting on the molecular degrees of
freedom are the same as the ground-state ones, $E_{\text{el}}\approx E_{0}+2E_{0}\Re\braket{\Delta u|u_{0}}\equiv E_{0}$,
and $F_{k}^{\text{NBO}}\approx2\Re\braket{\partial_{k}u_{0}|Q_{0}H'_{\text{el}}|\Delta u}$
represents the main effect of the electron reaction. Hence, upon plugging
the LRT result for $\ket{\Delta u}$ into the above expression for
$F_{k}^{\text{NBO}}$ one obtains a friction-like term $F_{k}^{\gamma}=-2\sum_{j}\Re\left(\int_{0}^{\infty}\Gamma_{kj}(\tau)V^{j}(t-\tau)d\tau\right)$,
where $V^{j}$ becomes the real particle velocity in the classical
limit, and $\Gamma$ is the kernel
\[
\Gamma_{kj}(\tau)=\braket{\partial_{k}u_{0}|Q_{0}H'_{\text{el}}e^{-\frac{i}{\hbar}H'_{\text{el}}\tau}|\partial_{j}u_{0}}
\]
In the Markov limit we can set $V^{j}(t-\tau)\approx V^{j}(t)$ and
the frictional force takes the form $F_{k}^{\gamma}=-\sum_{j}\bar{\gamma}_{kj}V^{j}$
where $\bar{\gamma}_{kj}$ is the zero-frequency limit of the frequency-dependent
kernel $\bar{\gamma}_{kj}(\omega)=2\lim_{\epsilon\rightarrow0^{+}}\int_{0}^{\infty}e^{-\epsilon\tau}e^{i\omega\tau}\Gamma_{kj}(\tau)d\tau$.
As shown in SM, the real part of the tensor $\bar{\gamma}_{kj}$ is
precisely the $T\rightarrow0$ limit of the mixed quantum-classical
DMS expression \cite{Dou2017}, Eq. \ref{eq:DMS friction}, and contains
a \emph{pseudo}-magnetic contribution $2\hbar\Im\braket{\partial_{k}u_{0}|Q_{0}|\partial_{j}u_{0}}=2\hbar\Im q_{kj}=-\hbar B_{kj}$.
The latter, when used to evaluate the forces acting on the nuclei,
is seen to give a term $+\hbar\sum_{j}B_{kj}\Re V^{j}$ that precisely
cancels the magnetic dynamical correction introduced above (see Eq.
\ref{eq:F_mag corrected}). This corrective effect is physically sound:
electronic friction cools the nuclear motion and enforces the adiabatic
limit, with its \emph{gauge} fields. Upon removing the spurious \emph{pseudo}-magnetic
contribution, the ordinary electronic friction tensor is the real
part of the expression 
\begin{equation}
\gamma_{kj}=2\pi\hbar\braket{\partial_{k}u_{0}|Q_{0}(H{}_{\text{el}}-E_{0})\delta(E_{0}^{+}-H{}_{\text{el}})|\partial_{j}u_{0}}\label{eq:Markov friction I}
\end{equation}
where $E_{0}^{+}=E_{0}+\hbar\omega$ in the limit $\omega\rightarrow0^{+}$
is understood. Notice the symmetries $\gamma'_{kj}=\gamma'_{jk}$
and $\gamma''_{kj}=-\gamma''_{jk}$ for the real and imaginary parts
of $\gamma_{kj}$, respectively, and the fact that $\gamma'_{kj}$
satisfies the second fluctuation-dissipation relation as described
in Ref. \cite{Dou2017} (in the $T\rightarrow0$ K limit considered
here), since the argument used by Dou \emph{et al.} remain valid in
this context. Furthermore, upon using $\bra{\partial_{k}u_{0}}Q_{0}H'_{\text{el}}=-\bra{u_{0}}(\partial_{k}H_{\text{el}})Q_{0}$
we can equivalently re-write $\gamma'_{kj}$ as 
\begin{equation}
\gamma'_{kj}=-2\pi\hbar\Re\braket{u_{0}|(\partial_{k}H_{\text{el}})\delta(E_{0}^{+}-H{}_{\text{el}})|\partial_{j}u_{0}}\label{eq:Markov friction II}
\end{equation}
It is this expression that reduces to Eq. \ref{eq:Head-Gordon Tully}
in the independent electron approximation (see SM), and that shows
more explicitly why $\gamma'_{kj}$ needs a manifold of states lying
infinitesimally close to the ground state to be non-vanishing. {[}The
required excitation energy $\hbar\omega\rightarrow0$ can also be
viewed as the ``running'' correction to $E_{0}$ in the dynamical
phase factor of the evolving electronic state.{]} 

In the context of the exact factorization of the wavefunction, the
electronic-friction regime considered here is best handled by using
local electronic states in the ``standard'' \emph{gauge,} \emph{i.e.,
}$\braket{u^{+}|\partial_{t}u^{+}}=0$ where $^{+}$ denotes this
\emph{gauge}. In the LRT limit this amounts to setting $E_{\text{el}}=E_{0}=0$
in the electronic problem and writing $\ket{u^{+}}=\ket{u_{0}}+\ket{\Delta u^{+}}$,
where now only the second term is time-dependent. This implies that
the nuclear Hamiltonian in the chosen \emph{gauge, $H^{+}$,} very
closely resembles the $n=0$ adiabatic Hamiltonian of Eq. \ref{eq:effective Hamiltonian}.
In fact, as shown in SM, the main difference is a modified vector
potential $A_{k}\rightarrow A_{k}+\delta A_{k}$ where 
\begin{align*}
\delta A_{k} & =2\Im\braket{\partial_{k}u_{0}|\Delta u_{0}}\\
 & \approx-2\sum_{j}\Im\int_{0}^{\infty}\braket{\partial_{k}u_{0}|e^{-\frac{i}{\hbar}H'_{\text{el}}t'}Q_{0}|\partial_{j}u_{0}}V_{t-t'}^{j}dt'
\end{align*}
This term is of second order in the spatial derivatives of the electronic
states and generates a correction of the same order to the adiabatic
forces through its time dependence, $F_{k}\rightarrow F_{k}-\hbar\partial_{t}\left(\delta A_{k}\right)$.
An integration by parts transforms $\delta A_{k}$ into 
\begin{equation}
\delta A_{k}=-2\sum_{j}\Im\left(q_{kj}X^{j}\right)+\frac{2}{\hbar}\sum_{j}\Re\left(\int_{0}^{\infty}\Gamma_{kj}(t')X_{t-t'}^{j}dt'\right)\label{eq:vector potential correction (general)}
\end{equation}
where $X^{j}$'s are the integrated velocity fields, $X^{j}(\mathbf{x},t)=\int_{-\infty}^{t}V^{j}(\mathbf{x},t')dt'$
and the first term is seen to be the source of the dynamical correction
to the \emph{pseudo}-Lorentz force (Eq.s \ref{eq:F_el corrected},\ref{eq:F_mag corrected}).
In the Markov limit, as seen above, only the integrated kernel $\bar{\gamma}_{kj}=2\lim_{\epsilon\rightarrow0^{+}}\int_{0}^{\infty}e^{-\epsilon\tau}\Gamma_{kj}(\tau)d\tau$
matters, for which we have $\bar{\gamma}_{kj}=-2i\hbar q_{kj}+\gamma_{kj}$.
Hence, in this limit Eq. \ref{eq:vector potential correction (general)}
reduces to
\begin{equation}
\delta A_{k}=\hbar^{-1}\sum_{j}\Re\left(\gamma_{kj}X_{t}^{j}\right)\label{eq:vector potential correction (Markov)}
\end{equation}
where the same cancellation effect occurs as discussed above.

Eq. (\ref{eq:vector potential correction (Markov)}) is a key result
of this Letter. It represents the simple amendment to the adiabatic
Hamiltonian that is necessary in order to include the effect of electronic
friction into the quantum dynamics of the nuclei, in the most relevant
case where the Markov limit applies. This term turns the Schr\"{o}dinger
equation into a non-linear equation of motion, which conserves the
wavefunction norm and describes energy dissipation. In the simplest
case where $\xi^{ij}=\delta^{ij}M^{-1}$, if $\gamma_{kj}$ can be
taken \emph{diagonal} and \emph{uniform} in the configuration space
of the system where the dynamics occurs, one finds 
\[
\delta A_{k}=\Re(\gamma X_{t}^{k})\approx\partial_{k}\Re\left(-i\hbar M^{-1}\gamma\int_{-\infty}^{t}\ln\psi_{t'}(\mathbf{x})dt'\right)
\]
upon neglecting the contribution of the vector potential to the velocity
field. Hence, $\delta A_{k}$ becomes longitudinal and can be replaced
by an appropriate \emph{scalar} field $\delta\phi\approx\hbar M^{-1}\gamma\Im\left(\ln\psi_{t}(\mathbf{x})\right)$
for negligible $\gamma''$. This scalar potential is precisely the
``phase potential'' appearing in the effective Hamiltonian of the
(non-linear) Kostin equation \cite{Kostin2003}, and represents a
very simple way to introduce dissipation into a Schr\"{o}dinger-like
equation. 

We remark that the resulting equation for the nuclear wavefunction
has no fluctuating term since it describes the evolution of electronically
averaged quantities, \emph{i.e.}, the nuclear observables dressed
by the electronic state. 

\textbf{\emph{Conclusions. }}We have developed a theory of electronic
friction that describes the nuclear dynamics in a quantum setting
at $T=0$ K. Friction is seen to turn the equation of motion for the
nuclear wavefunction into a non-linear equation, where the vector
potential depends on the past wavefunction behaviour. This low-temperature,
frictional limit seems to be ideal to explore the nuclear dynamics
in a situation where energy disspation through excitation of electron-hole
pairs of a metallic substrate \emph{enforces} the adiabatic dynamics.
We have shown that in this limit the \emph{gauge} fields appropriate
for an adiabatic dynamics are fully restored, hence we expect Berry's
phase effects to be operative. In molecular problems, the latter typically
reduce to a sign change, since the Berry's connection is flat in the
presence of time-reversal symmetry. However, in the presence of a
magnetic field (that here can come from the metallic surface itself)
a non-zero curvature is present, thereby allowing the geometric phase
to attain arbitrary values. A magnetic field also modifies the topology
of conical intersections by increasing the dimensionality of the branching
space, and turns the intersection points into \emph{pseudo}-magnetic
monopoles \cite{Berry1984}. More generally, the \emph{pseudo}-magnetic
field is known to shield the physical magnetic field and allow the
nuclei to behave essentially as neutral particles (as far as magnetic
effects are of concern) \cite{Resta2000}. However, the degree of
shielding depends on the nuclear geometry and may change along the
dynamical evolution \cite{Ceresoli2007}. The effects of these phenomena
on elementary gas-surface processes remain yet to be explored.\\

\bibliographystyle{apsrev4-1}

%
\pagebreak{}

\section*{Supplemental Material}

\subsection{\emph{Adiabatic approximation }}

\subsubsection*{I. Slow variables as external parameters}

When the slow variables $\mathbf{x}$ are regarded as parameters that
are under the control of the experimenter, the Hamiltonian governing
the evolution of the system has a pre-defined time-dependence $H=H(\mathbf{x}(t))$
for any given path $\mathbf{x}(t)$ of the parameters. The adiabatic
approximation can be recast as a variational approximation 
\[
i\hbar\frac{d}{dt}\ket{\Psi_{t}}=H(t)\ket{\Psi_{t}}\ \ \ket{\Psi_{0}}=\ket{u_{n}(0)}
\]
which uses a \emph{time-dependent} variational manifold $\mathcal{V}(t)$
(the $n^{\text{th}}$ eigenspace of $H(t)$, here assumed to be non-degenerate)
and a complex-analytic representation of the wavefunction. In other
words, one describes the wavefunction at any time in the form $\ket{\Psi}=C\ket{u_{n}(s)}$
where $C$ is the only (complex) \emph{variational} parameter of the
problem. The Dirac-Frenkel condition amounts to 
\[
P_{n}\left(i\hbar\partial_{t}-H(t)\right)\ket{\Psi_{t}}=0
\]
where $P_{n}=P_{n}(t)$ is the istantaneous eigenprojector on the
target manifold and $Q_{n}=1-P_{n}$ its orthogonal complement. Hence
\[
i\hbar\ket{\dot{\Psi}_{t}}=H_{PP}(t)\ket{\Psi_{t}}+i\hbar\dot{P}_{n}\ket{\Psi_{t}}
\]
where $Q_{n}\ket{\dot{\Psi}_{t}}\equiv\dot{P}_{n}\ket{\Psi_{t}}$
has been used to re-write the time-derivative in terms of the manifold
dynamics (contained in $\dot{P}_{n}$) and of $H_{PP}\equiv P_{n}HP_{n}$.
Furthermore, since $P_{n}\dot{P}_{n}\ket{\Psi_{t}}\equiv0$ the equation
of motion can be recast as an effective Schr\"{o}dinger equation
\[
i\hbar\frac{d}{dt}\ket{\Psi_{t}}=H_{n}(t)\ket{\Psi_{t}}
\]
involving the self-adjoint effective Hamiltonian 
\[
H_{n}(t)=H_{PP}(t)+i\hbar[\dot{P}_{n},P_{n}]
\]

\subsubsection*{Local-in-time-error in case I\protect \\
(Slow variables as parameters)}

The local-in-time error (LITE) accompanying a variational solution
\cite{Martinazzo2020} takes the general form 
\[
\varepsilon[\Psi]=\hbar^{-1}\left\Vert \left(i\hbar\partial_{t}-H\right)\Psi\right\Vert 
\]
and, in the case of a time-dependent, complex-analytic manifold, can
be given in terms of the above introduced variational Hamiltonian
as 
\begin{align*}
\hbar^{2}\varepsilon^{2}[\Psi] & =\left\Vert \left(H_{n}-H\right)\Psi\right\Vert ^{2}\\
 & \equiv\left\Vert \left(H-i\hbar\dot{P}_{n}\right)\Psi\right\Vert ^{2}-\left\Vert \left(H_{n}-i\hbar\dot{P}_{n}\right)\Psi\right\Vert ^{2}
\end{align*}
This result follows upon noticing that, for $\ket{\Psi}=P_{n}\ket{\Psi}$,
we have $\left(H-H_{n}\right)\ket{\Psi}=Q_{n}(H-i\hbar\dot{P}_{n})\ket{\Psi}$
hence 
\begin{align*}
\hbar^{2}\varepsilon^{2}[\Psi] & =\braket{\Psi|(H+i\hbar\dot{P}_{n})Q_{n}(H-i\hbar\dot{P}_{n})|\Psi}\\
 & =\braket{\Psi|(H+i\hbar\dot{P}_{n})(H-i\hbar\dot{P}_{n})|\Psi}+\\
 & -\braket{\Psi|(H+i\hbar\dot{P}_{n})P_{n}(H-i\hbar\dot{P}_{n})|\Psi}\\
 & \equiv||(H-i\hbar\dot{P}_{n})\Psi||^{2}-||H_{PP}\Psi||^{2}
\end{align*}
 which is the same as the above result since in the last expression
$H_{PP}$ can be replaced with $H_{n}-i\hbar\dot{P}_{n}$.

Now, upon noticing that in the adiabatic problem $H_{PP}\equiv E_{n}P_{n}$,
\[
\hbar^{2}\varepsilon^{2}[\Psi]\equiv\left\Vert E_{n}\Psi-i\hbar\dot{P}_{n}\Psi\right\Vert ^{2}-\left\Vert E_{n}\Psi\right\Vert ^{2}
\]
we find 
\[
\varepsilon=||\dot{P}_{n}\Psi||
\]
where $\braket{\Psi|\dot{P}_{n}|\Psi}=0$ has been used. This error
is a purely geometrical property, \emph{i.e.}, for a given infinitesimal
displacement in parameter space it does \emph{not} depend on the evolution
time. In fact, using $||\dot{P}_{n}\Psi||=||\dot{P}_{n}u_{n}||=\left\Vert Q_{n}\dot{u}_{n}\right\Vert $,
the LITE is seen to be a property of the moving manifold only and
takes the form $\varepsilon^{2}=\braket{\dot{u}_{n}|Q_{n}|\dot{u}_{n}}$
or, equivalently, introducing the parameter dependence,
\[
\varepsilon^{2}=\sum_{i,j}\braket{\partial_{i}u_{n}|Q_{n}|\partial_{j}u_{n}}\dot{x}^{i}\dot{x}^{j}=\sum_{ij}g_{ij}\dot{x}^{i}\dot{x}^{j}
\]
where $\partial_{i}\equiv\partial/\partial x^{i}$ and $g_{ij}$ is
the symmetric part of the covariant tensor on the parameter space
\[
q=\sum_{ij}\braket{\partial_{i}u_{n}|Q_{n}|\partial_{j}u_{n}}dx^{i}dx^{j}
\]
known as quantum geometric tensor. Such symmetric (real) part $g$
is a quantum metric (a Fubini-Study metric on the tangent bundle),
while the antisymmetric (imaginary) part is related to the Berry's
curvature. In a single equation, 
\[
q=g-\frac{i}{2}d\omega
\]
where $d$ denotes the exterior derivative and $\omega$ is the differential
form $\omega=\sum_{j}A_{j}dx^{j}$, i.e., 
\[
d\omega=\sum_{ij}B_{ij}dx^{i}dx^{j}
\]
Here, $A_{j}=i\braket{u_{n}|\partial_{j}u_{n}}$ subsumes the Berry's
connection on the vector bundle $\pi:E\rightarrow\mathcal{M}$ defined
by the parametric dependence of the given eigenspace and $B_{ij}\equiv\partial_{i}A_{j}-\partial_{j}A_{i}$
(here $\mathcal{M}$ is the nuclear configuration space and $\pi^{-1}(\text{x})$
is the $n^{\text{th}}$ eigenspace of the electronic Hamiltonian $H_{\text{el}}$). 

\subsubsection*{II. Slow variables as dynamical variables}

If the slow variables are considered as dynamical variables the electron-nuclear
wavefunction is written as 
\[
\ket{\Psi}=\int d\mathbf{x}\psi(\mathbf{x})\ket{u_{n}(\mathbf{x})}\ket{\mathbf{x}}
\]
The variational manifold is complex and application of the variational
principle reduces to the Dirac-Frenkel condition, 
\[
\int d\mathbf{x}'d\mathbf{x}\delta\psi^{*}(\mathbf{x}')\bra{u_{n}(\mathbf{x'}),\,\mathbf{x}'}\left[i\hbar\partial_{t}-H\right]\ket{u_{n}(\mathbf{x}),\,\mathbf{x}}\psi(\mathbf{x})=0
\]
Here 
\[
\braket{u_{n}(\mathbf{x'}),\,\mathbf{x}'|H|u_{n}(\mathbf{x}),\,\mathbf{x}}=\delta(\mathbf{x}-\mathbf{x}')\left(\braket{\hat{T}}_{n}+E_{n}(\mathbf{x})\right)
\]
where $E_{n}(\mathbf{x})$ is the Born - Oppenheimer potential energy
surface and $\braket{\hat{T}}_{n}$ is the coordinate-representation
of the nuclear kinetic energy operator \emph{averaged} over the electronic
state,
\[
\braket{\hat{T}}_{n}=\braket{u_{n}(\mathbf{x})|\hat{T}|u_{n}(\mathbf{x})}
\]
Setting $H_{n}=\braket{T}_{n}+E_{n}(\mathbf{x})$, the variational
equation of motion takes the form of a Schr\"{o}dinger equation for
the nuclear wavefunction
\[
H_{n}\psi=i\hbar\frac{\partial\psi}{\partial t}
\]
with an effective Hamiltonian specific of the electronic state under
consideration. The main difference with respect to the common Born-Oppenheimer
Hamiltonian lies in the nuclear kinetic operator which gets \emph{dressed}
by the electronic motion: this dressing is the way the \emph{gauge}
fields originate from the geometric properties of the adiabatic approximation.
In this context, it may be worth noticing that the derived equation
of motion, \emph{differently} \emph{from the BO Hamiltonian}, correctly
conserves energy since it is variational, and this occurs upon including
the above mentioned \emph{gauge} fields. 

Let us now examine the dressed kinetic energy operator. Let $T$ be
of the form 
\[
\hat{T}=\frac{1}{2}\sum_{ij}\xi^{ij}\hat{p}_{i}\hat{p}_{j}
\]
and notice that 
\begin{align*}
\braket{\hat{p}_{i}\hat{p}_{j}}_{n} & =\hat{p}_{i}\hat{p}_{j}-i\hbar\braket{u_{n}|\partial_{j}u_{n}}\hat{p}_{i}\\
 & -i\hbar\braket{u_{n}|\partial_{i}u_{n}}\hat{p}_{j}-\hbar^{2}\braket{u_{n}|\partial_{i}\partial_{j}u_{n}}
\end{align*}
where, as above, $A_{j}=i\braket{u_{n}|\partial_{j}u_{n}}\in\mathbb{R}$
and 
\begin{align*}
\braket{u_{n}|\partial_{i}\partial_{j}u_{n}} & =\partial_{i}\braket{u_{n}|\partial_{j}u_{n}}-\braket{\partial_{i}u_{n}|\partial_{j}u_{n}}\\
 & =-i\partial_{i}A_{j}-\braket{\partial_{i}u_{n}|Q\partial_{j}u_{n}}-\braket{\partial_{i}u_{n}|P|\partial_{j}u_{n}}\\
 & =-i\partial_{i}A_{j}-q_{ij}-A_{i}A_{j}
\end{align*}
where $q_{ij}$ is the $ij^{\text{th}}$ component of the quantum
geometric tensor (we omit the index $n$ from these quantities). Hence,
\[
\braket{\hat{p}_{i}\hat{p}_{j}}_{n}=\left(\hat{p}_{i}-\hbar A_{i}\right)\left(\hat{p}_{j}-\hbar A_{j}\right)+\hbar^{2}q_{ij}
\]
and, since $\xi^{ij}$ is symmetric,
\[
\braket{T}_{n}=\frac{1}{2}\sum_{ij}\xi^{ij}\left(\hat{p}_{i}-\hbar A_{i}\right)\left(\hat{p}_{j}-\hbar A_{j}\right)+\frac{\hbar^{2}}{2}\sum_{ij}\xi^{ij}g_{ij}
\]
Clearly, the \emph{dressed} operator contains terms analogous to a
vector ($A_{i}$) and scalar ($g_{ij}$) electromagnetic potential
and the latter modify the nuclear dynamics, when a comparison is made
with the simpler Born-Oppenheimer one. To see the effect of the \emph{gauge}
fields on the nuclear dynamics it is enough to consider the nuclear's
velocity 
\[
\hat{v}^{k}=\frac{i}{\hbar}[\braket{H}_{n},\hat{x}^{k}]=\sum_{j}\xi^{kj}\hat{\pi}_{j}
\]
where $\hat{\pi}_{k}=\hat{p}_{k}-\hbar A_{k}$ is thus the mechanical
momemtum for the $k^{\text{th}}$ degree of freedom. The latter satisfies
the commutation relation
\[
[\hat{\pi}_{i},\hat{\pi}_{j}]=i\hbar^{2}B_{ij}
\]
where $B_{ij}=\partial_{i}A_{j}-\partial_{j}A_{i}$ is the $ij^{\text{th}}$
component of the Berry's curvature, \emph{i.e.,} $d\omega\equiv\sum_{ij}B_{ij}dx^{i}dx^{j}$
(with $B_{ij}=-B_{ji}$). Notice that the commutator is \emph{gauge}
independent {[}The \emph{gauge} freedom mentioned here is the arbitrariness
in the choice of the electronic frame. A \emph{gauge} transformation
$\ket{u_{n}}\rightarrow e^{-i\varphi}\ket{u_{n}}$ amounts to adding
an \emph{exact} 1-form to $\omega$ without altering the scalar potential
($\omega\rightarrow\omega+d\varphi$) and, at the same time, to adding
a phase factor to the nuclear wavefunction, $\psi\rightarrow e^{+i\varphi}\psi$.
In this respect, the \emph{gauge} transformations involved here are
more limited than those allowed for a true electromagnetic potential,
since they are always stationary w.r.t. $t${]}. The force is then
obtained by the rate of variation of particles' mechanical momentum,
\[
\dot{\hat{\pi}}^{k}=\frac{i}{\hbar}[\braket{H}_{n},\hat{\pi}^{k}]=\frac{i}{\hbar}\left[\frac{1}{2}\sum_{ij}\xi^{ij}\hat{\pi}_{i}\hat{\pi}_{j}+E'_{n},\hat{\pi}^{k}\right]
\]
where $E'_{n}=E_{n}+\frac{\hbar^{2}}{2}\sum_{ij}\xi^{ij}g_{ij}$.
It is convenient to separate two contributions. The first does \emph{not}
involve derivatives of the inverse mass tensor and reads as 
\begin{align*}
\hat{W}^{k} & =\frac{i}{\hbar}\left\{ \frac{1}{2}\sum_{ij}\xi^{ij}[\hat{\pi}_{i}\hat{\pi}_{j},\hat{\pi}_{k}]+[E'_{n},\hat{\pi}_{k}]\right\} \\
 & =-\frac{\hbar}{2}\sum_{ij}\xi^{ij}\left(\hat{\pi}_{i}B_{jk}+B_{ik}\hat{\pi}_{j}\right)-\partial_{k}E'_{n}
\end{align*}
whereas the second only appears when the mass tensor is coordinate-dependent,
\begin{align*}
\hat{U}^{k} & =\frac{i}{2\hbar}\sum_{ij}[\xi^{ij},\hat{\pi}_{k}]\hat{\pi}_{i}\hat{\pi}_{j}=\frac{i}{2\hbar}\sum_{ij}[\xi^{ij},\hat{p}_{k}]\hat{\pi}_{i}\hat{\pi}_{j}\\
 & =-\frac{1}{2}\sum_{ij}\left(\partial_{k}\xi^{ij}\right)\hat{\pi}_{i}\hat{\pi}_{j}
\end{align*}

Henceforth we consider only the first term and write the $k^{\text{th}}$
component of the force as
\[
F_{k}=-\partial_{k}E'_{n}+\frac{\hbar}{2}\sum_{j}\left(\hat{v}^{j}B_{kj}+B_{kj}\hat{v}^{j}\right)
\]
that is, $F_{k}=F_{k}^{\text{BO}}+F_{k}^{\text{el}}+F_{k}^{\text{mag }}$,
where the first term represents the Born-Oppenheimer force 
\[
F_{k}^{\text{BO}}=-\frac{\partial E_{n}(\mathbf{x})}{\partial x^{k}}
\]
while the latter two form an effective Lorentz force which comprises
both an electric component 
\[
F_{k}^{\text{el}}=-\frac{\hbar^{2}}{2}\sum_{ij}\xi^{ij}\frac{\partial g_{ij}}{\partial x^{k}}
\]
arising from the Fubini-Study metric tensor, and a magnetic component
\[
F_{k}^{\text{mag}}=\frac{\hbar}{2}\sum_{j}\left(\hat{v}^{j}B_{kj}+B_{kj}\hat{v}^{j}\right)
\]
due to the Berry's curvature. (To check that this indeed represents
a \emph{pseudo}-magnetic force one can consider the three-dimensional
case and observe that $B_{xy}=H_{z}\ \ B_{xz}=-H_{y}\ \ B_{yx}=H_{x}$,
where $\mathbf{H}$ is the \emph{pseudo}-magnetic field. Hence $\mathbf{F}^{\text{mag}}=\frac{\hbar}{2}\left[\mathbf{v}\wedge\mathbf{H}-\mathbf{H}\wedge\mathbf{v}\right]$,
which is the correct quantum expression of the magnetic component
of the Lorentz force). The two components behave very differently
from each other: the \emph{pseudo}-magnetic field may vanish almost
everywhere (see below) yet give rise to observable effects, similarly
to what happens with the Aharonov-Bohm effect, while the \emph{pseudo}-electric
field is ubiquitous (\emph{i.e.}, it does not vanish unless the adiabatic
error is uniform over the configuration space sampled by the nuclei)
but typically of secondary importance and seldom considered in practice.
Amazingly, though, they both arise from one and the same object, namely
the quantum geometric tensor, here re-written in a form that makes
explicit its connections with the omitted adiabatic states 
\begin{align*}
q & =\sum_{ij}\sum_{m}\braket{\partial_{i}u_{n}|u_{m}}\braket{u_{m}|\partial_{j}u_{n}}dx^{i}dx^{j}\\
 & =\sum_{ij}\sum_{m}\frac{\braket{u_{n}|\partial_{i}H|u_{m}}\braket{u_{m}|\partial_{j}H|u_{n}}}{(E_{n}-E_{m})^{2}}dx^{i}dx^{j}
\end{align*}

\subsubsection*{Local-in-time error in case II\protect \\
{[}Slow variables as dynamical variables{]}}

Next we consider the LITE in the ``dynamic'' adiabatic approximation
\cite{Martinazzo2020}. To this end, we need the time-derivative of
the whole wavefunction in the ``standard'' \emph{gauge} $\braket{\Psi|\dot{\Psi}^{+}}=0$
(where the superscript $+$ is used for the trajectory $\ket{\Psi}=\ket{\Psi(t)}$
to denote this \emph{gauge}) and in particular its squared norm. By
this we mean
\[
\hbar^{2}||\dot{\Psi}^{+}||^{2}=\int d\mathbf{x}\psi^{*}(\mathbf{x})\left[\braket{\hat{T}}_{n}+(E_{n}(\mathbf{x})-\bar{E})\right]^{2}\psi(\mathbf{x})
\]
where $\braket{T}_{n}$ is the dressed kinetic energy operator introduced
above
\[
\braket{\hat{T}}_{n}=\frac{1}{2}\sum_{ij}\xi^{ij}\hat{\pi}_{i}\hat{\pi}_{j}+\phi
\]
with
\[
\hat{\pi}_{i}=\hat{p}_{i}-\hbar A_{i}\ \ \ \phi=\frac{\hbar^{2}}{2}\sum_{ij}\xi^{ij}g_{ij}
\]
and $\bar{E}=\braket{\Psi|H|\Psi}$ is the average total energy. We
also need the energy variance
\[
\Delta E^{2}=\int d\mathbf{x}\psi^{*}(\mathbf{x})\braket{\left[\hat{T}+(H_{\text{el}}(\mathbf{x})-\bar{E})\right]^{2}}_{n}\psi(\mathbf{x})
\]
and the result 
\begin{align*}
 & \braket{\left[\hat{T}+(H_{\text{el}}(\mathbf{x})-\bar{E})\right]^{2}}_{n}=\braket{\hat{T}^{2}}_{n}+\\
 & +\braket{(E_{n}(\mathbf{x})-\bar{E})^{2}}+2\Re\braket{\hat{T}(E_{n}(\mathbf{x})-\bar{E})}\\
 & =\braket{\hat{T}^{2}}_{n}+\\
 & +(E_{n}(\mathbf{x})-\bar{E})^{2}+2\Re\left[\braket{\hat{T}}_{n}(E_{n}(\mathbf{x})-\bar{E})\right]
\end{align*}
Hence, upon taking the difference of the two, we find for the local-in-time
error the expression
\[
\varepsilon^{2}[\Psi]=\frac{1}{\hbar^{2}}\int d\mathbf{x}\psi^{*}(\mathbf{x})\left[\braket{\hat{T}^{2}}_{n}-\braket{\hat{T}}_{n}^{2}\right]\psi(\mathbf{x})
\]
which shows explicitly the crucial role played by the nuclear kinetic
energy fluctuations in the adiabatic approximation.

This expression can also be put in a form that makes explicit the
contributions of electronic transitions. To this end it is worth introducing
the kinetic energy operator ``reduced'' with respect to the electronic
coordinates, $\braket{\hat{T}}_{nm}=\braket{u_{n}|\hat{T}|u_{m}}$
(the case $n=m$ reduces to the previous dressed kinetic energy operator
$\braket{T}_{n}$). These operators have a hermitian symmetry $\braket{T}_{nm}^{\dagger}=\braket{T}_{mn}$
(as can be readily checked by either their definition or a direct
calculation) and allow us to write 
\[
\braket{\hat{T}^{2}}_{nn}-\braket{\hat{T}}_{nn}^{2}=\sum_{m\neq n}\braket{\hat{T}}_{nm}\braket{\hat{T}}_{mn}\equiv\sum_{m\neq n}\braket{\hat{T}}_{mn}^{\dagger}\braket{\hat{T}}_{mn}
\]
In turn, upon defining
\[
\varphi_{m\leftarrow n}(\mathbf{x})=\braket{\hat{T}}_{mn}\psi(\mathbf{x})\equiv\braket{u_{m}|\hat{T}|u_{n}}_{\text{el}}\psi(\mathbf{x})
\]
we have the error in terms of contributing electronic transitions,
\[
\varepsilon_{\mathcal{V}}^{2}[\Psi]=\frac{1}{\hbar^{2}}\sum_{m\neq n}\int d\mathbf{x}|\varphi_{m\leftarrow n}(\mathbf{x})|^{2}
\]
where 
\begin{align*}
\nu_{m\leftarrow n}(\mathbf{x}) & =\frac{1}{\hbar^{2}}|\varphi_{m\leftarrow n}(\mathbf{x})|^{2}=\\
 & \frac{1}{\hbar^{2}}\psi^{*}(\mathbf{x})\bra{u_{n}}\hat{T}\ket{u_{m}}\bra{u_{m}}\hat{T}\ket{u_{n}}\psi(\mathbf{x})
\end{align*}
is a ``transition probability density'' which, in this form, is
manifestly \emph{gauge}-invariant since $\psi(\mathbf{x)}\ket{u_{n}}\equiv\braket{\mathbf{x}|\Psi}$. 

In order to make a closer comparison with the error obtained in the
previous section for the ``static'' adiabatic approximation, we
introduce $\nu(\mathbf{x})=\sum_{m\neq n}\nu_{m\leftarrow n}(\mathbf{x})$
and the total conditional transition probability density $\varrho(\mathbf{x})=\nu(\mathbf{x})/|\psi(\mathbf{x})|^{2}$
such that 
\[
\varepsilon^{2}[\Psi]=\int d\mathbf{x}|\psi(\mathbf{x})|^{2}\varrho(\mathbf{x})
\]
which is well defined provided $\psi(\mathbf{x})\neq0$. Clearly,
$\varrho(\mathbf{x})$ measures the error \emph{locally} in configuration
space (as well as in time), and thus describes the tendency of the
system in configuration $\mathbf{x}$ to jump to an electronic state
other than $n$. For a nuclear wavefunction $\psi(\mathbf{x})$ and
a frame $\ket{u_{n}(\mathbf{x})}$ in the vector bundle $\pi:E\rightarrow\mathcal{M}$
we consider $\ket{\psi_{n}(\mathbf{x})}=\psi(\mathbf{x})\ket{u_{n}(\mathbf{x})}$
as a smooth section of $E$, and the map to the normal bundle $\ket{\psi_{n}(\mathbf{x})}\rightarrow\ket{\varphi(\mathbf{x})}=Q\hat{T}\ket{\psi_{n}(\mathbf{x})}$
which gives $\nu(\mathbf{x})=\hbar^{-2}\braket{\varphi(\mathbf{x})|\varphi(\mathbf{x})}$.
We find
\begin{align*}
Q\hat{T}\ket{\psi_{n}(\mathbf{x})} & =\\
-\frac{\hbar^{2}}{2}\sum_{ij}\xi^{ij}\left[(\partial_{i}\psi)Q\ket{\partial_{j}u_{n}}\right.\\
\left.+(\partial_{j}\psi)Q\ket{\partial_{i}u_{n}}+\psi Q\ket{\partial_{i}\partial_{j}u_{n}}\right]
\end{align*}
Then, upon introducing $\hat{\pi}_{i}=-i\hbar\partial_{i}-\hbar A_{i}$,
\begin{align*}
Q\hat{T}\ket{\psi_{n}(\mathbf{x})} & =-i\hbar\sum_{ij}\xi^{ij}\left(\hat{\pi}_{i}\psi\right)Q\ket{\partial_{j}u_{n}}\\
 & -\frac{\hbar^{2}}{2}\sum_{ij}\xi^{ij}\psi D_{ij}\ket{u_{n}}\\
 & =-i\hbar\sum_{j}\left(\hat{v}^{j}\psi\right)Q\ket{\partial_{j}u_{n}}\\
 & -\frac{\hbar^{2}}{2}\psi\sum_{ij}\xi^{ij}D_{ij}\ket{u_{n}}
\end{align*}
where $\hat{v}^{j}$ is the $j^{\text{th}}$ component of the velocity
operator and
\[
D_{ij}\ket{u_{n}}=iA_{i}Q\ket{\partial_{j}u_{n}}+iA_{j}Q\ket{\partial_{i}u_{n}}+Q\ket{\partial_{i}\partial_{j}u_{n}}
\]
The reason why we introduce these two components is that they are
separately \emph{gauge}-invariant: under the \emph{gauge} transformation
$\ket{u_{n}}\rightarrow\ket{u_{n}}e^{-i\varphi}$, $\psi\rightarrow\psi e^{+i\varphi}$
we have $A_{i}\rightarrow A_{i}+\partial_{i}\varphi$ and 
\begin{align*}
 & Q\ket{\partial_{j}u_{n}}\rightarrow e^{-i\varphi}Q\ket{\partial_{j}u_{n}}\\
 & \hat{\pi}_{i}\psi\rightarrow e^{i\varphi}\hat{\pi}_{i}\psi\\
 & D_{ij}\ket{u_{n}}=e^{-i\varphi}D_{ij}\ket{u_{n}}
\end{align*}
since (there is no need to verify it with an explicit calculation,
since both $\ket{\psi_{n}(\mathbf{x})}$ and the velocity term are
\emph{gauge} invariant) 
\begin{align*}
Q\ket{\partial_{i}\partial_{j}u_{n}} & \rightarrow\\
e^{-i\varphi}\left[Q\ket{\partial_{i}\partial_{j}u_{n}}-i(\partial_{i}\varphi)Q\ket{\partial_{j}u_{n}}-i(\partial_{j}\varphi)Q\ket{\partial_{i}u_{n}}\right]
\end{align*}
Stated differently, the operators $Q\partial_{j},\hat{\pi}_{i},\hat{v}^{i},D_{ij},$
etc. are \emph{tensorial} under \emph{gauge} transformations. Hence,
upon introducing the (\emph{gauge-}tensorial) residue 
\[
R\ket{u_{n}}=\frac{\hbar}{2}\sum_{ij}\xi^{ij}D_{ij}\ket{u_{n}}
\]
and the (complex-valued)\textbf{ }quantum velocity fields $V^{j}$
\[
V^{j}=\frac{\hat{v}^{j}\psi}{\psi}=\frac{\psi^{*}\hat{v}^{j}\psi}{|\psi|^{2}}\equiv V^{j}(\mathbf{x})
\]
we find
\begin{align*}
\varrho(\mathbf{x}) & =\sum_{ij}\left(V^{i}\right)^{*}V^{j}q_{ij}\\
 & -i\sum_{j}\left(V^{j}\right)^{*}\braket{\partial_{j}u_{n}|Ru_{n}}+\\
 & +i\sum_{j}V^{j}\braket{Ru_{n}|\partial_{j}u_{n}}\\
 & +\braket{Ru_{n}|Ru_{n}}
\end{align*}
where $q_{ij}=\braket{\partial_{i}u_{n}|Q|\partial_{j}u_{n}}$ is
the quantum geometric tensor and the remaining scalar products contain
higher derivatives of the electronic state in a \emph{gauge} invariant
form. The first term closely resembles the local-in-time error in
the standard adiabatic approximation analyzed in the previous section
\[
\varepsilon^{2}=\sum_{ij}V^{i}V^{j}q_{ij}
\]
where $V^{i}$ is now the classical velocity of the $i^{\text{th}}$
parameter. There are though notable differences: when turning the
slow variables into quantum variables \emph{both} the real (symmetric)
\emph{and} the imaginary (antisymmetric) parts of $q_{ij}$ matter
for the error, since
\[
\sum_{ij}\left(V^{i}\right)^{*}V^{j}q_{ij}=\sum_{ij}K^{ij}g_{ij}+\frac{1}{2}\sum_{ij}Y^{ij}B_{ij}
\]
where 
\[
K^{ij}=\Re\left(\left(V^{i}\right)^{*}\left(V^{j}\right)\right)\ \ Y^{ij}=\Im\left(\left(V^{i}\right)^{*}\left(V^{j}\right)\right)
\]
and $g_{ij}$ and $B_{ij}$ have been introduced above. It is instructive
then to consider their total contribution upon integrating over configuration
space. For the first we find
\begin{align*}
\int d\mathbf{x}|\psi(\mathbf{x})|^{2}\sum_{ij}K^{ij}g_{ij} & =\sum_{ij}\int d\mathbf{x}\Re\left(\left(\hat{v}^{i}\psi\right)^{*}g_{ij}\left(\hat{v}^{j}\psi\right)\right)\\
 & =\Re\left(\bra{\psi}\sum_{ij}\hat{v}^{i}g_{ij}\hat{v}^{j}\ket{\psi}_{X}\right)\\
 & \equiv\bra{\psi}\sum_{ij}\hat{v}^{i}g_{ij}\hat{v}^{j}\ket{\psi}_{X}
\end{align*}
where the scalar product $\braket{.|.}_{X}$ is that of the Hilbert
space $L^{2}(\mathcal{M})$ describing the nuclear degrees of freedom
and where the last equality follows from the fact that the operator
$\sum_{ij}\hat{v}^{i}g_{ij}\hat{v}^{j}$ is self-adjoint on that space.
As for the second we have similarly 
\begin{align*}
\int d\mathbf{x}|\psi(\mathbf{x})|^{2}\sum_{ij}Y^{ij}B_{ij} & =\sum_{ij}\int d\mathbf{x}\Im\left(\left(\hat{v}^{i}\psi\right)^{*}B_{ij}\left(\hat{v}^{j}\psi\right)\right)\\
 & =\Im\left(\bra{\psi}\sum_{ij}\hat{v}^{i}B_{ij}\hat{v}^{j}\ket{\psi}_{X}\right)\\
 & =-i\bra{\psi}\sum_{ij}\hat{v}^{i}B_{ij}\hat{v}^{j}\ket{\psi}_{X}
\end{align*}
since the operator $\sum_{ij}\hat{v}^{i}B_{ij}\hat{v}^{j}$ is anti-hermitian
\[
\left(\sum_{ij}\hat{v}^{i}B_{ij}\hat{v}^{j}\right)^{\dagger}=\sum_{ij}\hat{v}^{j}B_{ij}\hat{v}^{i}=-\sum_{ij}\hat{v}^{i}B_{ij}\hat{v}^{j}
\]
Hence, overall, by considering the ``classical'' contribution only,
we find that the LITE in the dynamic adiabatic approximation is just
the expectation value of a self-adjoint quantum tensor\textbf{
\[
\hat{q}=\sum_{ij}\hat{v}^{i}\left(g_{ij}-\frac{i}{2}B_{ij}\right)\hat{v}^{j}\equiv\sum_{ij}\hat{v}^{i}q_{ij}\hat{v}^{j}
\]
}which is nothing but the quantum version of the quantum geometric
tensor\textbf{.} That is, to leading order, we have
\[
\varepsilon^{2}\approx\bra{\psi}\sum_{ij}\hat{v}^{i}q_{ij}\hat{v}^{j}\ket{\psi}_{X}
\]
On comparing with the static adiabatic approximation, however, one
should also observe that additional terms appear whose physical meaning
is far less obvious. 

\subsubsection*{Non-adiabatic transition probability}

To understand better the meaning of $\varrho(\mathbf{x})$ and its
components $\varrho_{m\leftarrow n}$ we consider the situation in
which, during the time evolution, the local-in-time error exceeds
a given threshold, thereby suggesting the need of going \emph{beyond}
the adiabatic approximation. This can be accomplished dynamically
by ``spawning''\cite{Martinazzo2020} the electronic basis that
forms the variational manifold, \emph{e.}g., by expanding the wavefunction
\emph{ansatz }to
\[
\ket{\Psi_{t}}=\int d\mathbf{x}\psi_{t}(\mathbf{x})\ket{u_{n}(\mathbf{x}),\mathbf{x}}+\int d\mathbf{x}\phi_{t}(\mathbf{x})\ket{u_{s}(\mathbf{x}),\mathbf{x}}
\]
where $s=n\pm1$ depending on which gap $|E_{n}-E_{n\pm1}|$ is the
smallest. Henceforth, we shall first address the simpler situation
where a single neighboring state affects the dynamics and later generalize
the result to a multitude of electronic states. 

At the time of spawning $t_{s}$ the amplitude $\phi_{t}(\mathbf{x})$
\emph{must} vanish and its time derivative is determined by the variational
equations of motion
\[
\left\{ \begin{array}{c}
i\hbar\frac{\partial\psi}{\partial t}=\left(\braket{T}_{nn}+E_{n}\right)\psi+\braket{T}_{ns}\phi\\
\\
i\hbar\frac{\partial\phi}{\partial t}=\braket{T}_{sn}\psi+\left(\braket{T}_{ss}+E_{s}\right)\phi
\end{array}\right.
\]
which give, for $t=t_{s}$, 
\[
i\hbar\frac{\partial\phi}{\partial t}\bigg|_{t_{s}}=\braket{T}_{sn}\psi\equiv\varphi_{s\leftarrow n}
\]
(notice that the \emph{gauge} does not affect this off-diagonal term).
Thus, we see that the probability 
\[
\nu_{s\leftarrow n}=\hbar^{-2}\int d\mathbf{x}|\varphi_{s\leftarrow n}|^{2}
\]
represents precisely the \emph{error} \emph{reduction} \emph{due to
electronic spawning},
\[
\varepsilon_{\mathcal{V}}^{2}\rightarrow\varepsilon_{\mathcal{V}'}^{2}=\varepsilon_{\mathcal{V}}^{2}-\nu_{s\leftarrow n}\ \ \ \text{at}\ t=t_{s}
\]
\emph{i.e.,} the error reduction arising from lifting the adiabatic
approximation by allowing non-adiabatic transitions to the state $s$.
On the other hand, the above equation also determines the short-time
behaviour of the non-adiabatic transition probability $P_{s}$ to
the state $s$ as 
\[
P_{s}\approx\nu_{s\leftarrow n}(t-t_{s})^{2}\ \ t\geq t_{s}
\]
since
\[
\frac{d|\phi|^{2}}{dt}\bigg|_{t_{s}}=\phi^{*}\frac{d\phi}{dt}+\frac{d\phi^{*}}{dt}\phi\bigg|_{t_{s}}\equiv0
\]
and 
\[
\frac{d^{2}|\phi|^{2}}{dt^{2}}\bigg|_{t_{s}}=2\frac{d\phi^{*}}{dt}\frac{d\phi}{dt}\bigg|_{t_{s}}\equiv\frac{2}{\hbar^{2}}\left|\braket{T}_{sn}\psi\right|^{2}
\]
{[}Notice that $\nu_{s\leftarrow n}$ is half the second derivative
of the transition probability at $t=t_{s}$, a result which follows
in general from the definition of local-in-time error.{]} This finding
leads to an interesting conclusion: when a single term $s$ dominates
the sum, $\nu_{s\leftarrow n}$ is approximately the total squared
error in the dynamic adiabatic approximation and we have seen above
that this is determined by the quantum geometric tensor (to leading
order in the derivatives of the $\ket{u_{n}}$'s).\textbf{ }Hence,
turning this argument around, we see that the quantum geometric tensor
also determines the early transition probability upon spawning. In
other words, we have approximately, up to second order in $\delta t=t-t_{s}$,
\begin{align*}
P_{s} & \approx\int d\mathbf{x}\sum_{ij}\left(\delta\hat{x}^{i}\psi_{t_{s}}\right)^{*}(\mathbf{x})\left(\delta\hat{x}^{j}\psi_{t_{s}}\right)(\mathbf{x})q_{ij}(\mathbf{x})\\
 & \ \ \text{with}\ \delta\hat{x}^{i}:=\hat{v}^{i}\delta t
\end{align*}
if the most important non-adiabatic channel were suddenly opened at
time $t_{s}$. 

More generally, all the above remains unaltered if the adiabatic approximation
is suddenly lifted and the ``spawning'' process is made virtually
complete, \emph{i.e.}, the variational constraint is suddenly removed
at $t=t_{s}$ and the wavefunction is allowed to expand into the whole
Hilbert space 
\begin{align*}
\ket{\Psi_{t}} & =\int d\mathbf{x}\psi_{t}(\mathbf{x})\ket{u_{n}(\mathbf{x}),\mathbf{x}}\rightarrow\\
 & \ket{\Psi_{t}}=\int d\mathbf{x}\psi_{t}(\mathbf{x})\ket{u_{n}(\mathbf{x}),\mathbf{x}}\\
 & +\sum_{m\neq n}\int d\mathbf{x}\phi_{t}^{(m)}(\mathbf{x})\ket{u_{m}(\mathbf{x}),\mathbf{x}}
\end{align*}
Again, we have $\phi^{(m)}\equiv0$ at the time of spawning, 
\[
i\hbar\frac{\partial\phi^{(m)}}{\partial t}\bigg|_{t_{s}}=\braket{T}_{mn}\psi\equiv\varphi_{m\leftarrow n}
\]
holds for any $m\neq n$ and now the local-in-time error is reduced
\emph{exactly} to zero upon spawning. Thus, the \emph{total} non-adiabatic
transition probability $P$ can be given, up to second order in $\delta t$,
as
\begin{align*}
P & \approx\int d\mathbf{x}\sum_{ij}\left(\delta\hat{x}^{i}\psi_{t_{s}}\right)^{*}(\mathbf{x})\left(\delta\hat{x}^{j}\psi_{t_{s}}\right)(\mathbf{x})q_{ij}(\mathbf{x})\\
 & \ \ \text{with}\ \delta\hat{x}^{i}:=\hat{v}^{i}\delta t
\end{align*}
under the only assumption that the terms involving the second derivatives
of the electronic states are negligible. This result relates the geometric
properties of the adiabatic problem to the rate of non-adiabatic transitions.
In a sense, this is an obvious result since the latter transitions
represent precisely the failure of the adiabatic approximation. At
a closer look, though, it is rather surprising that the exact dynamics
of the system beyond the adiabatic paradigm is determined solely by
the geometric properties of the approximation.

\subsection{\emph{Exact factorization of the wavefunction }}

As mentioned in the main text, the exact factorization \cite{Abedi2010,Abedi2012}
is an ``intermediate'' representation which is obtained by introducing
a local basis of nuclear states $\{\ket{\mathbf{x}}\}$ to represent
the exact wavefunction describing the combined electron-nuclear states,
\emph{i.e.}, 
\[
\ket{\Psi_{t}}=\int d\mathbf{x}\ket{\mathbf{x}}\braket{\mathbf{x}|\Psi_{t}}
\]
where $\braket{\mathbf{x}|\Psi_{t}}$\emph{ }is yet a vector in the
electronic Hilbert space $\mathcal{H}_{\text{el}}$ that we write
as \emph{
\[
\braket{\mathbf{x}|\Psi_{t}}=\psi_{t}(\mathbf{x})\ket{u_{t}(\mathbf{x})}
\]
}upon imposing a normalization condition and choosing a smoothly varying
phase for the \emph{local} electronic states $\ket{u_{t}(\mathbf{x})}$.
This gives the wavefunction in the (local) exactly-factorized representation
\[
\ket{\Psi_{t}}=\int d\mathbf{x}\psi_{t}(\mathbf{x})\ket{\mathbf{x}}\ket{u_{t}(\mathbf{x})}
\]
Clearly, there is some freedom in choosing $\ket{u_{t}}$ (and correspondingly
in defining the nuclear wavefunction $\psi_{t}$) that we may fix
by imposing the arbitrary (but real) \emph{gauge} term
\[
A_{0}=i\braket{u|\partial_{t}u}
\]
in the equation of motion, besides the usual Berry's connection terms
$A_{k}=i\braket{u|\partial_{k}u}$.

\subsubsection*{Equations of motion}

To obtain the equations of motion for the above nuclear wavefunction
and the electronic state we write the total Hamiltonian using the
coordinate representation for the nuclear variables, \emph{i.e.} in
the form 
\[
\hat{H}=\hat{T}+H_{\text{el}}(\mathbf{x})
\]
where $\hat{T}$ is the nuclear kinetic energy operator
\[
\hat{T}=\frac{1}{2}\sum_{ij}\xi^{ij}\hat{p}_{i}\hat{p}_{j}\ \ \ \text{with}\ \hat{p}_{j}=-i\hbar\partial_{j}
\]
and $H_{\text{el}}(\mathbf{x})$ is the electronic operator with the
nuclei clamped at a configuration $\mathbf{x}$. 

From the Schr\"{o}dinger equation 
\[
\hat{H}(\psi\ket{u})=i\hbar\left(\partial_{t}\psi\right)\ket{u}+i\hbar\psi\ket{\partial_{t}u}
\]
we immediately obtain the equation of motion for the nuclear wavefunction
by projecting the above equation onto $\ket{u}$
\[
\left(\braket{\hat{H}}_{\text{el}}-\hbar A_{0}\right)\psi=i\hbar\left(\partial_{t}\psi\right)
\]
where $\braket{\hat{H}}_{\text{el}}=\braket{\hat{T}}_{\text{el}}+\braket{u|H_{\text{el}}|u}$
contains the dressed kinetic energy operator and the ``Born-Oppenheimer''
average energy $E_{\text{el}}=\braket{u|H_{\text{el}}|u}$, and $A_{0}$
is the \emph{gauge} potential introduced above. We remark that $\braket{\hat{T}}_{\text{el}}$
appearing here is the nuclear kinetic energy operator \emph{averaged}
over the time-dependent electronic state and reads explicitly
\[
\braket{\hat{T}}_{\text{el}}=\frac{1}{2}\sum_{ij}\xi^{ij}\hat{\pi}_{i}\hat{\pi}_{j}+\frac{1}{2}\sum_{ij}\xi^{ij}q_{ij}
\]
where $\hat{\pi}_{j}=\hat{p}_{j}-\hbar A_{j}$ and $q_{ij}=\braket{\partial_{i}u|Q|\partial_{j}u}$.
We use the hat symbol to remind us the coordinate representation adopted,
but notice that $\ket{u}$ is everywhere meant to be the time-dependent
electronic state (correspondingly, $P=\ket{u}\bra{u},$ $Q=1-P$,
etc.).

As for the equation governing the electron dynamics we only need its
projection onto the ``unoccupied'' electronic space, since $P\partial_{t}\ket{u}=\ket{u}\braket{u|\partial_{t}u}$
is known once the \emph{gauge} term $A_{0}$ has been fixed. Hence,
\[
Q\hat{H}(\psi\ket{u})=+i\hbar\psi Q\ket{\partial_{t}u}
\]
which gives 
\[
i\hbar Q\ket{\partial_{t}u}=\frac{1}{\psi}Q\hat{H}\left(\psi\ket{u}\right)
\]
or, if we write the equation for $\partial_{t}\ket{u}$, 
\[
i\hbar\ket{\partial_{t}u}=+\hbar A_{0}\ket{u}+\frac{1}{\psi}Q\hat{H}\left(\psi\ket{u}\right)
\]
Here, the effective electronic Hamiltonian operator contains two terms
\[
\frac{1}{\psi}Q\hat{H}\left(\psi\ket{u}\right)=\frac{1}{\psi}Q\hat{T}\left(\psi\ket{u}\right)+QH_{\text{el}}\ket{u}
\]
but only the first depends on $\psi$ since $H_{\text{el}}$ is local
in nuclear coordinates. The first term, which we denote as $K[\psi]\ket{u}$,
is found to be
\[
K[\psi]\ket{u}=-i\hbar\sum_{j}V^{j}Q\ket{\partial_{j}u}-\hbar R\ket{u}
\]
where $V^{j}=(\hat{v}^{j}\psi)/\psi$ is the complex-valued nuclear
velocity field, $R\ket{u}=\frac{\hbar}{2}\sum_{ij}\xi^{ij}D_{ij}\ket{u}$,
and
\[
D_{ij}\ket{u}=iA_{i}Q\ket{\partial_{j}u}+iA_{j}Q\ket{\partial_{i}u}+Q\ket{\partial_{i}\partial_{j}u}
\]
We stress that the above decomposition has simple \emph{gauge} transformation
properties, since $V^{j}$ is \emph{gauge} invariant and both $Q\partial_{j}$
and $D_{ij}$ (hence $R$) behave tensorially under a \emph{gauge}
transformation. Hence,
\[
i\hbar Q\ket{\partial_{t}u}=QH_{\text{el}}+K[\psi]\ket{u}
\]
where, on the r.h.s., the first term describes the electron dynamics
with the nuclei clamped at $\mathbf{x}$ and the second term describes
the drag effect on the electrons due to the motion of the nuclei. 

\subsubsection*{Equivalence with the formuation of Abedi et al.}

The above equations for the nuclear and electronic ``wavefunctions''
are identical to those given in Ref. \cite{Abedi2010,Abedi2012}.
This is evident for the nuclear equation but not for the electronic
equation since the authors of Ref. \cite{Abedi2010,Abedi2012} wrote
it in a rather different form which, in our notation, would read
\begin{align*}
i\hbar\ket{\partial_{t}u} & =\left[H_{\text{el}}-(\bar{E}-\hbar A_{0}+\frac{\hbar^{2}}{2}\sum_{ij}\xi^{ij}q_{ij})\right]\ket{u}+\\
 & +\left[\sum_{ij}\frac{\xi^{ij}}{2}\left(\hat{p}_{i}-\hbar A_{i}\right)\left(\hat{p}_{j}+\hbar A_{j}\right)+\right.\\
 & \left.+\sum_{ij}\xi^{ij}\left(\frac{\hat{p}_{i}\psi}{\psi}\right)\left(\hat{p}_{j}+\hbar A_{j}\right)\right]\ket{u}
\end{align*}
Here, $(\bar{E}-\hbar A_{0}+\frac{\hbar^{2}}{2}\sum_{ij}\xi^{ij}q_{ij})=\varepsilon$
is the effective energy introduced by the authors of Ref. \cite{Abedi2010,Abedi2012}
and the second bracket, denoted $F$ in the following, contains Hamiltonian
momentum terms $\hat{p}_{i}$'s rather than $\hat{\pi}_{i}$'s or
$\hat{v}^{i}$'s (which are \emph{gauge} tensorial). However, it is
only a matter of simple algebra to show that indeed 
\[
F-\varepsilon=K[\psi]+\hbar A_{0}-\bar{E}
\]
as required by the equation above or, equivalently, 
\[
F=K[\psi]+\frac{\hbar^{2}}{2}\sum_{ij}\xi^{ij}q_{ij}
\]
To see this notice that 
\begin{align*}
F & \equiv\sum_{ij}\xi^{ij}\left(\frac{\hat{\pi}_{i}}{2}+\frac{\hat{\pi}_{i}\psi}{\psi}+\hbar A_{i}\right)\left(\hat{p}_{j}+\hbar A_{j}\right)\\
 & =\sum_{ij}\xi^{ij}\left(\frac{\hat{p}_{i}+\hbar A_{i}}{2}+\frac{\hat{\pi}_{i}\psi}{\psi}\right)\left(\hat{p}_{j}+\hbar A_{j}\right)\\
 & =\sum_{ij}\frac{\xi^{ij}}{2}\hat{p}_{i}\hat{p}_{j}+\sum_{ij}\frac{\xi^{ij}}{2}\hbar\hat{p}_{i}A_{j}+\sum_{ij}\frac{\xi^{ij}}{2}\hbar A_{i}\hat{p}_{j}\\
 & +\frac{\hbar^{2}}{2}\sum_{ij}\xi^{ij}A_{i}A_{j}+\sum_{ij}\frac{\hat{v}^{j}\psi}{\psi}\left(\hat{p}_{j}+\hbar A_{j}\right)
\end{align*}
where 
\[
\left(\hat{p}_{j}+\hbar A_{j}\right)\ket{u}=-i\hbar\left(\partial_{j}-\braket{u|\partial_{j}u}\right)\ket{u}\equiv-i\hbar Q\partial_{j}\ket{u}
\]
 gives 
\[
\sum_{j}\frac{\hat{v}^{j}\psi}{\psi}\left(\hat{p}_{j}+\hbar A_{j}\right)\ket{u}=-i\hbar\sum_{j}\frac{\hat{v}^{j}\psi}{\psi}Q\partial_{j}\ket{u}
\]
and, on the other hand, 
\begin{align*}
 & \sum_{ij}\frac{\xi^{ij}}{2}\hbar(\hat{p}_{i}A_{j}+A_{i}\hat{p}_{j})\ket{u}=\\
 & -i\hbar^{2}\sum_{ij}\frac{\xi^{ij}}{2}(\partial_{i}A_{j})\ket{u}+\\
 & -i\hbar^{2}\sum_{ij}\frac{\xi^{ij}}{2}(A_{i}\ket{\partial_{j}u}+A_{j}\ket{\partial_{i}u})=\\
 & =-i\hbar^{2}\sum_{ij}\frac{\xi^{ij}}{2}(A_{i}Q\ket{\partial_{j}u}+A_{j}Q\ket{\partial_{i}u})+\\
 & -i\hbar^{2}\sum_{ij}\frac{\xi^{ij}}{2}(\partial_{i}A_{j})\ket{u}-\hbar^{2}\sum_{ij}\xi^{ij}A_{i}A_{j}\ket{u}
\end{align*}
Hence,
\begin{align*}
F\ket{u} & =-i\hbar\sum_{j}\frac{\hat{v}^{j}\psi}{\psi}Q\partial_{j}\ket{u}\\
 & -i\hbar^{2}\sum_{ij}\frac{\xi^{ij}}{2}(A_{i}Q\ket{\partial_{j}u}+A_{j}Q\ket{\partial_{i}u})\\
 & +\sum_{ij}\frac{\xi^{ij}}{2}\hat{p}_{i}\hat{p}_{j}-i\hbar^{2}\sum_{ij}\frac{\xi^{ij}}{2}(\partial_{i}A_{j})\ket{u}\\
 & -\frac{\hbar^{2}}{2}\sum_{ij}\xi^{ij}A_{i}A_{j}\ket{u}
\end{align*}
Finally, upon observing that 
\begin{align*}
A_{i}A_{j}+q_{ij} & =\braket{\partial_{i}u|u}\braket{u|\partial_{j}u}+\braket{\partial_{i}u|Q\partial_{j}u}\\
 & \equiv\braket{\partial_{i}u|\partial_{j}u}\equiv\partial_{i}\left(\braket{u|\partial_{j}u}\right)-\braket{u|\partial_{i}\partial_{j}u}
\end{align*}
we write 
\begin{align*}
-\frac{\hbar^{2}}{2}\sum_{ij}\xi^{ij}A_{i}A_{j}\ket{u}-i\hbar^{2}\sum_{ij}\frac{\xi^{ij}}{2}\left(\partial_{i}A_{j}\right)\ket{u}\\
=\frac{\hbar^{2}}{2}\sum_{ij}\xi^{ij}P\ket{\partial_{i}\partial_{j}u}+\frac{\hbar^{2}}{2}\sum_{ij}\xi^{ij}q_{ij}\ket{u}
\end{align*}
and obtain
\begin{align*}
F\ket{u} & =-i\hbar\sum_{j}\frac{\hat{v}^{j}\psi}{\psi}Q\partial_{j}\ket{u}\\
 & -\frac{\hbar^{2}}{2}\sum_{ij}\xi^{ij}\left(iA_{i}Q\ket{\partial_{j}u}+iA_{j}Q\ket{\partial_{i}u}+Q\ket{\partial_{i}\partial_{j}u}\right)\\
 & +\frac{\hbar^{2}}{2}\sum_{ij}\xi^{ij}q_{ij}\ket{u}
\end{align*}
where 
\[
iA_{i}Q\ket{\partial_{j}u}+iA_{j}Q\ket{\partial_{i}u}+Q\ket{\partial_{i}\partial_{j}u}\equiv D_{ij}\ket{u}
\]
\emph{i.e.}, 
\[
F\ket{u}=-i\hbar\sum_{j}\frac{\hat{v}^{j}\psi}{\psi}Q\partial_{j}\ket{u}-\hbar R\ket{u}+\frac{\hbar^{2}}{2}\sum_{ij}\xi^{ij}q_{ij}\ket{u}
\]
as we intended to show.

\subsubsection*{Dynamically corrected pseudo-Lorentz force: \protect \\
proof of the vanishing of its average}

In the main text, we have mentioned that introducing the time derivative
of the electronic state 
\[
\hbar Q\partial_{t}\ket{u}=-iQH_{\text{el}}\ket{u}-iK[\psi_{t}]\ket{u}
\]
in the electron dynamical force 
\[
F_{k}^{\text{ED}}=-2\Im\braket{\partial_{k}u|Q|\hbar\partial_{t}u}
\]
one obtains a genuine non-adiabatic term
\[
F_{k}^{\text{nad}}=2\Re\braket{\partial_{k}u|QH_{\text{el}}|u}
\]
and a correction 
\[
F_{k}^{\text{corr}}=2\Re\braket{\partial_{k}u|K[\psi_{t}]|u}
\]
to the \emph{pseudo}-Lorentz force that makes the latter vanish on
average. We give here the details of the calculation, starting from
the observation that, with $-iK[\psi_{t}]\ket{u}=-\hbar\sum_{j}V^{j}Q\ket{\partial_{j}u}+i\hbar R\ket{u}$,
we obtain
\begin{align*}
F_{k}^{\text{corr}} & =2\hbar\sum_{j}\Im(\braket{\partial_{k}u|Q\partial_{j}u}V^{j})-2\hbar\Re\braket{\partial_{k}u|Ru}\\
 & \equiv2\hbar\sum_{j}g_{kj}\Im V^{j}-\hbar\sum_{j}B_{kj}\Re V^{j}-2\hbar\Re\braket{\partial_{k}u|Ru}
\end{align*}
where $\Im q_{kj}=-B_{kj}/2$ has been used. The corrected magnetic
force is easily identified 
\[
F_{k}^{\text{mag,c}}=\frac{\hbar}{2}\sum_{j}\left(\hat{v}^{j}B_{kj}+B_{kj}\hat{v}^{j}\right)-\hbar\sum_{j}B_{kj}\Re V^{j}
\]
and found to have zero average with a simple calculation,
\begin{align*}
\braket{\psi|F_{k}^{\text{mag,c}}|\psi}_{X} & =\frac{\hbar}{2}\sum_{j}\braket{\psi|\hat{v}^{j}B_{kj}+B_{kj}\hat{v}^{j}|\psi}_{\text{X}}\\
 & -\hbar\sum_{j}\braket{\psi|B_{kj}\Re V^{j}|\psi}_{X}
\end{align*}
since 
\[
\frac{1}{2}\braket{\psi|\hat{v}^{j}B_{kj}+B_{kj}\hat{v}^{j}|\psi}_{\text{X}}=\Re\braket{\psi|B_{kj}\hat{v}^{j}|\psi}_{\text{X}}
\]
and, on the other hand, 
\begin{align*}
 & \braket{\psi|B_{kj}\Re V^{j}|\psi}_{X}=\\
 & =\int d\mathbf{x}\psi^{*}(\mathbf{x})B_{kj}(\mathbf{x})\frac{\Re\left(\psi(\mathbf{x})^{*}\hat{v}^{j}\psi(\mathbf{x})\right)}{|\psi(\mathbf{x})|^{2}}\psi(\mathbf{x})\\
 & =\int d\mathbf{x}B_{kj}(\mathbf{x})\Re\left(\psi(\mathbf{x})^{*}\hat{v}^{j}\psi(\mathbf{x})\right)\\
 & =\Re\int d\mathbf{x}B_{kj}(\mathbf{x})\psi(\mathbf{x})^{*}\hat{v}^{j}\psi(\mathbf{x})\\
 & =\Re\braket{\psi|B_{kj}\hat{v}^{j}|\psi}_{\text{X}}
\end{align*}
Hence, 
\[
\braket{\psi|F_{k}^{\text{mag,c}}|\psi}_{X}\equiv0
\]
for any state of the nuclei. The corrected \emph{pseudo-}electric
force reads as 
\begin{align*}
F_{k}^{\text{el,c}} & =2\hbar\sum_{j}g_{kj}\Im V^{j}\\
 & -\hbar^{2}\sum_{ij}\xi^{ij}\Re\braket{\partial_{k}u|D_{ij}u}-\frac{\hbar^{2}}{2}\sum_{ij}\xi^{ij}\frac{\partial g_{ij}}{\partial x^{k}}
\end{align*}
and can be re-written in a more symmetric form 
\begin{align*}
F_{k}^{\text{el,c}} & =2\hbar\sum_{j}g_{kj}\Im V^{j}\\
 & -\hbar^{2}\sum_{ij}\xi^{ij}\left(\Re\braket{\partial_{i}u|D_{kj}u}+\Re\braket{\partial_{k}u|D_{ij}u}\right)
\end{align*}
This can be seen by observing that 
\begin{align*}
\frac{\partial g_{ij}}{\partial x^{k}} & \equiv\frac{\partial}{\partial x^{k}}\Re\braket{\partial_{i}u|QQ\partial_{j}u}\\
 & =\Re\braket{\partial_{i}u|Q\frac{\partial}{\partial x^{k}}Q\partial_{j}u}+\Re\braket{\partial_{j}u|Q\frac{\partial}{\partial x^{k}}Q\partial_{i}u}
\end{align*}
where the (\emph{gauge}-invariant) derivatives 
\begin{align*}
Q\frac{\partial}{\partial x^{k}}Q\ket{\partial_{j}u} & =-Q\left(\partial_{k}P\right)\ket{\partial_{j}u}+Q\ket{\partial_{k}\partial_{i}u}\\
 & \equiv+iA_{j}Q\ket{\partial_{k}u}+Q\ket{\partial_{k}\partial_{j}u}\\
 & \equiv D_{kj}\ket{u}-iA_{k}Q\ket{\partial_{j}u}
\end{align*}
can be used to write 
\begin{align*}
\frac{\partial g_{ij}}{\partial x^{k}} & =\Re\braket{\partial_{i}u|D_{kj}u}+A_{k}\Im q_{ij}+\Re\braket{\partial_{j}u|D_{ki}u}+A_{k}\Im q_{ji}\\
 & \equiv\Re\braket{\partial_{i}u|D_{kj}u}+\Re\braket{\partial_{j}u|D_{ki}u}
\end{align*}
(this is of course symmetric w.r.t. exchange of $i$ and $j$). Now,
on taking the average
\begin{align*}
\bra{\psi}2\hbar\sum_{j}g_{kj}\Im V^{j}\ket{\psi}_{X} & =2\hbar\sum_{j}\Im\int d\mathbf{x}\psi^{*}(\mathbf{x})g_{kj}(\mathbf{x})\hat{v}^{j}\psi(\mathbf{x})\\
 & \equiv2\hbar\sum_{j}\braket{\psi|\Im(g_{kj}\hat{v}^{j})|\psi}
\end{align*}
where 
\begin{align*}
2\hbar\Im(g_{kj}\hat{v}^{j}) & =-i\hbar[g_{kj},\hat{v}^{j}]=\\
-i\hbar\sum_{i}\xi^{ij}[g_{kj},\hat{\pi}_{i}]= & -i\hbar\sum_{i}\xi^{ij}[g_{kj},\hat{p}_{i}]\equiv\hbar^{2}\sum_{i}\xi^{ij}\frac{\partial g_{kj}}{\partial x^{i}}
\end{align*}
hence 
\begin{align*}
\braket{\psi|F_{k}^{\text{el,c}}|\psi} & =\hbar^{2}\sum_{ij}\xi^{ij}\bra{\psi}\\
 & \left[\frac{\partial g_{ki}}{\partial x^{j}}-\left(\Re\braket{\partial_{i}u|D_{kj}u}+\Re\braket{\partial_{k}u|D_{ij}u}\right)\right]\ket{\psi}
\end{align*}
where the operator to be averaged reads as 
\begin{align*}
\Re\braket{\partial_{k}u|D_{ij}u}+\Re\braket{\partial_{i}u|D_{kj}u}+\\
-\Re\braket{\partial_{j}u|D_{ki}u}-\Re\braket{\partial_{k}u|D_{ij}u} & \equiv0
\end{align*}
since $\Re\braket{\partial_{i}u|D_{kj}u}=\Re\braket{\partial_{j}u|D_{ki}u}$.
In fact, importantly, we have exploited 
\[
\text{\ensuremath{\Re}}\braket{\partial_{k}u|D_{ij}u}=\frac{1}{2}\left(\frac{\partial g_{ik}}{\partial x^{j}}+\frac{\partial g_{kj}}{\partial x^{i}}-\frac{\partial g_{ij}}{\partial x^{k}}\right)
\]
that shows how $\text{\ensuremath{\Re}}\braket{\partial_{k}u|D_{ij}u}$
is related to the connection $\nabla^{q}$ on the tangent bundle induced
by the Fubini-Study metric, 
\[
\text{\ensuremath{\Re}}\braket{\partial_{k}u|D_{ij}u}=g_{kl}\Gamma_{ij}^{l}
\]
where $\Gamma_{ij}^{l}$ is the Christoffel symbol of the connection.
Indeed, the zeroing of the average \emph{pseudo}-electric force merely
expresses the conservation of the metric by the corresponding Ricci-Levi
Civita connection 
\begin{align*}
\braket{\psi|F_{k}^{\text{el,c}}|\psi} & =\hbar^{2}\sum_{ij}\xi^{ij}\braket{\psi|\frac{\partial g_{ki}}{\partial x^{j}}-g_{il}\Gamma_{kj}^{l}-g_{kl}\Gamma_{ij}^{l}|\psi}\\
 & =\hbar^{2}\sum_{ij}\xi^{ij}\braket{\psi|\left(\nabla_{j}^{q}g\right)_{ik}|\psi}
\end{align*}
where $\left(\nabla_{j}^{q}g\right)_{ik}$ is the $ik^{\text{th}}$
component of the covariant derivative of $g$ taken with the connection
$\nabla^{q}$ along the direction $j$. 

\subsubsection*{Statistical properties}

We emphasize here that, despite its role of a marginal probability
amplitude, \textbf{$\psi(\mathbf{x})$ }alone cannot determine the
full statistical properties of the nuclear subset of particles, not
even instantaneously, \emph{i.e.}, at a given instant of time. This
is evident from the fact that the (instantaneous) statistical properties
require the reduced density operator $\rho_{X}=\text{tr}_{e}\rho$
which for pure states and the factorization introduced above reads
\begin{align*}
\braket{\mathbf{x}|\rho_{X}|\mathbf{x'}} & =\text{tr}_{e}\left(\psi(\mathbf{x})\ket{u(\mathbf{x})}\bra{u(\mathbf{x'})}\psi^{*}(\mathbf{x}')\right)\\
 & =\sigma(\mathbf{x},\mathbf{x}')\braket{u(\mathbf{x}')|u(\mathbf{x})}
\end{align*}
where $\sigma(\mathbf{x},\mathbf{x}')=\psi(\mathbf{x})\psi^{*}(\mathbf{x'})$
is the ``apparent'' nuclear density matrix. In view of this, we
have two different strategies (and interpretative tools) to investigate
the statitistical properties of nuclear observables. Either we use
the true density matrix $\rho_{X}(\mathbf{x},\mathbf{x}')$ and bare
nuclear observables $N$
\[
\braket{N}=\int d\mathbf{x}\int d\mathbf{x}'\rho_{X}(\mathbf{x},\mathbf{x}')N(\mathbf{x'},\mathbf{x})
\]
or we use the\textbf{ }apparent density matrix $\sigma(\mathbf{x},\mathbf{x}')$
and dressed nuclear observables $\tilde{N}$, 
\[
\braket{N}=\int d\mathbf{x}\int d\mathbf{x}'\sigma(\mathbf{x},\mathbf{x}')\tilde{N}(\mathbf{x'},\mathbf{x})
\]
where 
\[
\tilde{N}(\mathbf{x},\mathbf{x}')=N(\mathbf{x},\mathbf{x}')\braket{u(\mathbf{x})|u(\mathbf{x}')}
\]
or, equivalently, 
\[
\tilde{N}(\mathbf{x},\mathbf{x}')=\braket{u(\mathbf{x})|N(\mathbf{x},\mathbf{x}')|u(\mathbf{x}')}_{\text{el}}
\]
which shows that the dressed observables are ``averaged'' over the
electronic states. 

As for the electronic density operator $\rho_{\text{el}}$, it takes
the form of a convex combination of electronic density operators $\rho_{\text{el}}(\mathbf{x})$
\[
\rho_{\text{el}}=\int_{X}d\mathbf{x}P(\mathbf{x})\rho_{\text{el}}(\mathbf{x})
\]
where $P(\mathbf{x})=|\psi(\mathbf{x})|^{2}$ is the probability density
of finding the nuclei at $\mathbf{x}$ and $\rho_{\text{el}}(\mathbf{x})$
is the conditional density operator
\[
\rho_{\text{el}}(\mathbf{x})=\frac{\braket{\mathbf{x}|\rho|\mathbf{x}}}{P(\mathbf{x})}\equiv\ket{u(\mathbf{x})}\bra{u(\mathbf{x})}
\]
which describes a pure local state, the one defined locally by the
exact factorization representation. 

For comparison, notice that the results in the adiabatic approximation
are very similar to the one given here, the only difference being
that $\ket{u}$ is replaced by a stationary state. Therefore, the
concept that the adiabatic approximation ``artificially'' forces
the \emph{local} electronic state to be a pure state is misleading,
because this is true for an arbitrary wavefunction.

In order to clarify the meaning of observables dressed by the electronic
state let us consider in detail the nuclear momentum for the $k^{\text{th}}$
nuclear degree of freedom, $\hat{p}_{k}$ (in the coordinate representation
appropriate for the exact facorization). This is first ``extended''
to an operator $\hat{P}_{k}=\hat{p}_{k}\otimes\mathbb{I}_{\text{el}}$
acting on the Hilbert space of the electronic-nuclear system, and
then ``reduced'' to an operator $\tilde{p}_{k}$ on the nuclear
space by averaging over the electronic state
\[
\tilde{p}_{k}=\braket{u|\hat{P}_{k}|u}=\hat{p}_{k}-i\hbar\braket{u|\partial_{k}u}\equiv\hat{\pi}_{k}
\]
The result is the operator for the mechanical momentum $\hat{\pi}_{k}$
introduced in the main text, which can thus be considered the canonical
momentum dressed by the electronic state. In general, for notational
convenience, one does not distinguish $\hat{P}_{k}$ from $\hat{p}_{k}$,
and then care is needed in interpreting $\hat{p}_{k}$ as the ``microscopic''
operator acting on the electronic-nuclear space or the ``averaged''
one acting on the nuclear space only. As for the dressed operators,
they are always averaged over the electronic state, and thus meant
to be operators on the Hilbert space of the nuclei.

It is instructive at this point to re-consider the total force $F_{k}$
acting on the $k^{\text{th}}$ nuclear degree of freedom in light
of the above difference between ``microscopic'' and ``electronically
averaged'' quantities. On the one hand we have
\begin{align*}
\frac{d\braket{\hat{p}_{k}}}{dt} & =\braket{\Psi|\frac{i}{\hbar}[H,\hat{p}_{k}]|\Psi}\\
 & =\braket{\Psi|-\partial_{k}H_{\text{el}}|\Psi}\\
 & =\int d\mathbf{x}\psi^{*}(\mathbf{x})\braket{-\partial_{k}H_{\text{el}}}_{\text{el}}\psi(\mathbf{x})
\end{align*}
where, to avoid confusion, we used the subscript el on the angular
bracket to denote the electronic average. This shows that the average
total force is the expectation value of the dressed microscopic force
$-\partial_{k}H_{\text{el}}$ acting on the given nuclear degree of
freedom. The latter can be equivalently re-written as
\begin{align*}
\braket{-\partial_{k}H_{\text{el}}}_{\text{el}} & =F_{k}^{\text{BO}}+2\Re\braket{\partial_{k}u|H_{\text{el}}|u}\\
 & \equiv F_{k}^{\text{BO}}+F_{k}^{\text{NBO}}
\end{align*}
since $\Re\left(\braket{\partial_{k}u|u}\braket{u|H_{\text{el}}|u}\right)=0$,
in order to make evident the Born-Oppenheimer-like contribution. On
the other hand, we also have
\[
\frac{d\braket{\hat{p}_{k}}}{dt}=2\Re\braket{\Psi|\hat{p}_{k}|\partial_{t}\Psi}
\]
where the time-derivative of the total wavefunction in the exact factorized
form can be written as
\[
\partial_{t}(\psi\ket{u})=\left[(\partial_{t}\psi)+\psi\braket{u|\partial_{t}u}\right]\ket{u}+\psi Q\ket{\partial_{t}u}
\]
Here, the term between square brackets amounts to 
\[
\left[(\partial_{t}\psi)+\psi\braket{u|\partial_{t}u}\right]=-\frac{i}{\hbar}\braket{H}_{\text{el}}\psi
\]
and thus 
\begin{align*}
\frac{d\braket{\hat{p}_{k}}}{dt} & =2\Re\int d\mathbf{x}\psi^{*}(\mathbf{x})\left(-\frac{i}{\hbar}\right)\left(\hat{\pi}_{k}\braket{H}_{\text{el}}\right)\psi(\mathbf{x})+\\
 & +2\Re\int d\mathbf{x}\psi^{*}(\mathbf{x})\left[+i\hbar\braket{\partial_{k}u|Q\partial_{t}u}\right]\psi(\mathbf{x})
\end{align*}
Here, for the first line we have used $\braket{\hat{p}_{k}}_{\text{el}}=\hat{\pi}_{k}$,
whereas for the second one we have exploited 
\begin{align*}
\braket{u|\hat{p}_{k}Q\partial_{t}u} & =-i\hbar\bra{u}\left(\partial_{k}Q\ket{\partial_{t}u}\right)+\braket{u|Q\partial_{t}u}\hat{p}_{k}\\
 & =-i\hbar\partial_{t}\left(\braket{u|Q\partial_{t}u}\right)+i\hbar\braket{\partial_{k}u|Q\partial_{t}u}\\
 & \equiv i\hbar\braket{\partial_{k}u|Q\partial_{t}u}
\end{align*}
since $\braket{u|Q\partial_{t}u}\equiv0$ (here the scalar products
are all meant to be on the electronic space only). Furthermore, since
\begin{align*}
2\Re\int d\mathbf{x}\psi^{*}(\mathbf{x})\left(-\frac{i}{\hbar}\right)\left(\hat{\pi}_{k}\braket{H}_{\text{el}}\right)\psi(\mathbf{x}) & =\\
=\int d\mathbf{x}\psi^{*}(\mathbf{x})\frac{i}{\hbar}\left[\braket{H}_{\text{el}},\hat{\pi}_{k}\right]\psi(\mathbf{x})
\end{align*}
and the second term is the expecation value of $F_{k}^{\text{ED}}$
introduced in the main text, we finally arrive at
\[
\frac{d\braket{\hat{p}_{k}}}{dt}=\int d\mathbf{x}\psi^{*}(\mathbf{x})\left[F_{k}^{\text{BO}}+F_{\text{k}}^{\text{mag}}+F_{k}^{\text{el}}+F_{k}^{\text{ED}}\right]\psi(\mathbf{x})
\]
On comparing with the previous expression and remembering that $F_{\text{k}}^{\text{mag}}+F_{k}^{\text{el}}+F_{k}^{ED}=F_{\text{k}}^{\text{mag,c}}+F_{k}^{\text{el,c}}+F_{k}^{\text{ED}}$
we find,
\[
\int d\mathbf{x}\psi^{*}(\mathbf{x})\left[F_{\text{k}}^{\text{mag,c}}+F_{k}^{\text{el,c}}\right]\psi(\mathbf{x})=0
\]
This is consistent with the result given in the previous section,
however the proof given there makes clear that the dynamically corrected
\emph{pseudo}-electric and \emph{pseudo}-magnetic forces vanish separately
when averaged. 

\subsection{Electronic friction}

\subsubsection*{Linear response}

Let us consider the integral form of the electronic equation in the
spirit of linear response theory, set $\hbar A_{0}\equiv E_{\text{el}}=\braket{u|H_{\text{el}}|u}$
and assume that $\ket{u(t_{0})}=e^{-\frac{i}{\hbar}E_{0}t_{0}}\ket{u_{0}}$
holds for some initial time $t_{0}$ in the infinite past. Let us
first take the simpler non-adiabatic term represented by the following
impulsive `kick' 
\[
\delta(t-\tau)\tilde{K}[\psi_{\tau}]\ket{u(\tau)}
\]
that acts instantanesouly, \emph{i.e.,} only at time $t=\tau$. The
electronic state soon after the kick reads
\[
\ket{u(\tau^{+})}\approx\ket{u(\tau^{-})}-\frac{i}{\hbar}K_{0}[\psi_{\tau}]\ket{u(\tau^{-})}
\]
where $\ket{u(\tau^{-}}=e^{-\frac{i}{\hbar}E_{0}\tau}\ket{u_{0}}$
is the freely propagating state and $Q\rightarrow Q_{0}=1-\ket{u_{0}}\bra{u_{0}}$
has been used for $t=\tau-\epsilon$, $\epsilon>0$. This follows
from the integral version of the equation of motion by shrinking the
time interval around the kick time $\tau$. Note that, correspondingly,
$K$ has been replaced by $K_{0}$ to remind us the use of $Q_{0}$
rather than $Q$, and of the ground-state connection in the velocity
operators. Hence, for any time $t$, we have
\begin{align*}
\ket{u(t)} & \approx e^{-\frac{i}{\hbar}E_{0}t}\ket{u_{0}}\\
 & -\frac{i}{\hbar}e^{-\frac{i}{\hbar}H_{\text{el}}(t-\tau)}K_{0}[\psi_{\tau}]\ket{u_{0}}e^{-\frac{i}{\hbar}E_{0}\tau}\Theta(t-\tau)
\end{align*}
where $\Theta(t)=1$ for $t>0$ and zero otherwise. Now, when considering
the full driving term
\[
K[\psi_{t}]\ket{u(t)}=\int_{-\infty}^{+\infty}d\tau\delta(t-\tau)K[\psi_{\tau}]\ket{u(\tau)}
\]
we have, assuming linear response, 
\begin{align*}
\ket{u(t)} & \approx e^{-\frac{i}{\hbar}E_{0}t}\ket{u_{0}}\\
 & -\frac{i}{\hbar}\int_{-\infty}^{+\infty}e^{-\frac{i}{\hbar}H_{\text{el}}(t-\tau)}K_{0}[\psi_{\tau}]\ket{u_{0}}e^{-\frac{i}{\hbar}E_{0}\tau}\Theta(t-\tau)d\tau
\end{align*}
hence 
\[
\ket{\Delta u}\approx-\frac{i}{\hbar}e^{-\frac{i}{\hbar}E_{0}t}\int_{0}^{\infty}e^{-\frac{i}{\hbar}(H_{\text{el}}-E_{0})t'}K_{0}[\psi_{t-t'}]\ket{u_{0}}dt'
\]
is such that $\braket{\Delta u|u_{0}}=0$ since $K_{0}=Q_{0}K_{0}$. 

Plugging this expression in the genuine non-Born-Oppenheimer force
given above we obtain two terms, 
\[
F_{k}^{\text{NBO,I}}=-2\sum_{j}\Re\int_{0}^{\infty}\Gamma_{kj}(\tau)V^{j}(t-\tau)d\tau
\]
with the kernel
\[
\Gamma_{kj}(t)=\braket{\partial_{k}u_{0}|Q_{0}H'_{\text{el}}e^{-\frac{i}{\hbar}H'_{\text{el}}t}|\partial_{j}u_{0}}
\]
and 
\[
F_{k}^{\text{NBO,II}}=-2\Im\int_{0}^{\infty}\braket{\partial_{k}u_{0}|Q_{0}H'_{\text{el}}e^{-\frac{i}{\hbar}H'_{\text{el}}\tau}R|u_{0}}d\tau
\]
where we have set $H'_{\text{el}}=H_{\text{el}}-E_{0}.$ The first
is a friction-like term, with $2\Re\Gamma_{kj}(t)$ playing the role
of memory kernel in the classical limit where $V^{j}$ is real. In
the Markov limit 
\[
F_{k}^{\text{NBO,I}}=-\sum_{j}\bar{\gamma}_{kj}V^{j}(t)
\]
where 
\[
\bar{\gamma}_{kj}=2\lim_{\epsilon\rightarrow0^{+}}\int_{0}^{\infty}e^{-\epsilon t}\Gamma_{kj}(t)dt
\]
with the usual $\epsilon$ converging factor included. Later we shall
find that $\bar{\gamma}_{kj}$ is better defined as the zero-frequency
limit (from above) of the frequency-dependent kernel 
\[
\bar{\gamma}_{kj}(\omega)=2\lim_{\epsilon\rightarrow0^{+}}\int_{0}^{\infty}e^{-\epsilon t}e^{i\omega t}\Gamma_{kj}(t)dt
\]
where the excitation energy $\hbar\omega$ can be viewed as a ``running''
correction to $E_{0}$ in the dynamical phase factor $e^{\frac{i}{\hbar}E_{0}t}$
appearing in $\Gamma_{kj}(t)$. 

\subsubsection*{Equivalence with DMS friction at T=0 K}

Let us first prove the equivalence of the DMS expression for the friction
(Eq. \ref{eq:DMS friction} of the main text) with the one obtained
above. When the electronic bath is not carrying any current the steady-state
density operator is the canonical one, and in the limit $T\rightarrow0$
we have $\rho\rightarrow\ket{u_{0}}\bra{u_{0}}=P_{0}$ and $\partial_{j}\rho=\ket{\partial_{j}u_{0}}\bra{u_{0}}+\ket{u_{0}}\bra{\partial_{j}u_{0}}$.
This gives two terms
\begin{align*}
\gamma_{kj}^{\text{DMS}} & =-\int_{0}^{\infty}\text{tr}_{e}\left(\left(\partial_{k}H_{\text{el}}\right)e^{-\frac{i}{\hbar}H'{}_{\text{el}}\tau}\ket{\partial_{j}u_{0}}\bra{u_{0}}\right)d\tau\\
 & -\int_{0}^{\infty}\text{tr}_{e}\left(\left(\partial_{k}H_{\text{el}}\right)\ket{u_{0}}\bra{\partial_{j}u_{0}}e^{+\frac{i}{\hbar}H'{}_{\text{el}}\tau}\right)d\tau
\end{align*}
which are the complex conjugate of each other, \emph{i.e.}, 
\[
\gamma_{kj}^{\text{DMS}}=-2\Re\int_{0}^{\infty}\braket{u_{0}|\left(\partial_{k}H_{\text{el}}\right)e^{-\frac{i}{\hbar}H'{}_{\text{el}}\tau}|\partial_{j}u_{0}}d\tau
\]
Then, upon noticing that 
\[
(\partial_{k}H)\ket{u_{0}}=(E_{0}-H_{\text{el}})\ket{\partial_{k}u_{0}}+(\partial_{k}E_{0})\ket{u_{0}}
\]
and introducing the projector $Q_{0}=1-P_{0}$ we find 
\begin{align*}
\gamma_{kj}^{\text{DMS}} & =2\Re\int_{0}^{\infty}\braket{\partial_{k}u_{0}|Q_{0}H'_{\text{el}}e^{-\frac{i}{\hbar}H'{}_{\text{el}}\tau}|\partial_{j}u_{0}}d\tau\\
 & -2(\partial_{k}E_{0})\int_{0}^{\infty}\Re\braket{u_{0}|\partial_{j}u_{0}}d\tau
\end{align*}
where the first term is precisely the real part of $\bar{\gamma}_{kj}$
introduced above, and the second term vanishes identically since $\braket{u_{0}|\partial_{j}u_{0}}$
is pure imaginary. Note that the usual converging factor has been
here tacitly assumed.

\subsubsection*{Pseudo-magnetic contribution}

Secondly, we show that the memoryless friction 
\[
\bar{\gamma}_{kj}=2\lim_{\epsilon\rightarrow0^{+}}\int_{0}^{\infty}e^{-\epsilon t}\Gamma_{kj}(t)dt
\]
contains in fact a \emph{pseudo}-magnetic contribution. To this end
we need 
\[
H'_{\text{el}}e^{-\epsilon t}e^{-\frac{i}{\hbar}H'_{\text{el}}t}=i\hbar\frac{d}{dt}\left(e^{-\epsilon t}e^{-\frac{i}{\hbar}H'_{\text{el}}t}\right)+i\hbar\epsilon\left(e^{-\epsilon t}e^{-\frac{i}{\hbar}H'_{\text{el}}t}\right)
\]
and
\begin{align*}
H'_{\text{el}}\int_{0}^{\infty}e^{-\epsilon t}e^{-\frac{i}{\hbar}H'_{\text{el}}t}dt & =-i\hbar\left(1+\frac{i\epsilon\hbar}{H'_{\text{el}}-i\epsilon\hbar}\right)\\
 & =-i\hbar\left(1+i\epsilon\hbar\frac{H'_{\text{el}}+i\epsilon\hbar}{(H'_{el})^{2}+\epsilon^{2}\hbar^{2}}\right)\\
 & \rightarrow-i\hbar\left(1+i\pi H'_{\text{el}}\delta(H'_{\text{el}})\right)\\
\end{align*}
where we have used the common notation $\frac{1}{A}$ for $A^{-1}$.
Notice that in this expression the second term on the r.h.s. $\propto H'_{\text{el}}\delta(H'_{\text{el}})$
would vanish if it were applied to a regular electronic state, but
this is not the case here because of the presence of the derivative
couplings. 

Plugging this identity in the friction expression we find 
\begin{align*}
\Re\bar{\gamma}_{kj} & =\gamma_{kj}^{\text{DMS}}=2\hbar\Im\braket{\partial_{k}u_{0}|Q_{0}|\partial_{j}u_{0}}\\
 & +2\pi\hbar\Re\braket{\partial_{k}u_{0}|Q_{0}H'_{\text{el}}\delta(H'_{\text{el}})|\partial_{j}u_{0}}
\end{align*}
where the first term
\[
2\hbar\Im\braket{\partial_{k}u_{0}|Q_{0}|\partial_{j}u_{0}}=2\hbar\Im q_{kj}=-\hbar B_{kj}
\]
gives a magnetic component $+\hbar\sum_{j}B_{kl}V^{j}$ that, when
evaluating the force, precisely cancels the magnetic correction introduced
in the main text. This term does not appear in the common case when
the electronic states can be taken as real functions of the electron
coordinates (as DMS assumed), which is possible in the absence of
magnetic fields and for a trivial topology of the ground adiabatic
state. It is however necessary when the magnetic field is turned on
or if conical intersections exist that can be encircled by the evolving
nuclear wavepacket. This corrective effect may thus be viewed physically
as restoration of the full adiabatic dynamics in this friction limit:
electronic friction cools the nuclear motion and enforces the adiabatic
limit (with its \emph{gauge} fields). 

To summarize
\[
\bar{\gamma}_{kj}=-2i\hbar q_{kj}+\gamma_{kj}
\]
where the ``corrected'' friction kernel $\gamma_{kj}$ takes the
form of the real part of the expression
\[
\gamma_{kj}=2\pi\hbar\braket{\partial_{k}u_{0}|Q_{0}H'_{\text{el}}\delta(H'_{\text{el}})|\partial_{j}u_{0}}
\]
or, equivalently,
\[
\gamma_{kj}=2i\hbar\lim_{\epsilon\rightarrow0^{+}}\epsilon\int_{0}^{\infty}e^{-\epsilon t}\braket{\partial_{k}u_{0}|Q_{0}e^{-\frac{i}{\hbar}H'_{\text{el}}t}|\partial_{j}u_{0}}
\]

\subsubsection*{Running correction to $E_{0}$}

The friction kernel $\gamma_{kj}$ defined above is, strictly speaking,
ill-defined, as it is apparent from the presence of both $Q_{0}$
and $\delta(H'_{\text{el}})$. A more appropriate definition is obtained
by taking the zero-frequency limit (from above) of the frequency-dependent
kernel 
\[
\gamma_{kj}(\omega)=2i\hbar\lim_{\epsilon\rightarrow0^{+}}\epsilon\int_{0}^{\infty}e^{-\epsilon t}e^{i\omega t}\braket{\partial_{k}u_{0}|Q_{0}e^{-\frac{i}{\hbar}H'_{\text{el}}t}|\partial_{j}u_{0}}
\]
The physical motivation for introducing a small (eventually vanishing
in the end) positive frequency $\omega$ is that the evolving ground-electronic
state has an energy slightly above $E_{0}$, \emph{i.e.} $E_{0}+\hbar\omega$
for $\hbar\omega\rightarrow0^{+}$, right because of excitations of
$e-h$ pairs into the substrate. Thus the replacement
\[
\exp\left(\frac{i}{\hbar}E_{0}t\right)\rightarrow\exp(i\omega t)\exp\left(\frac{i}{\hbar}E_{0}t\right)
\]
is needed to correct the LRT result for this effect. 

Before addressing this issue in detail, let us first derive some relationships
needed to handle the derivative couplings, and useful to derive different
equivalent expressions for the friction kernel. Let first $E_{0}$
be a discrete, non-degenerate energy eigenvalue of the electronic
Hamiltonian $H_{\text{el}}$ for some value of the nuclear coordinates
$\mathbf{x}=(x^{1},x^{2},..x^{k},..)$. Upon taking the derivative
of the electronic Schr\"{o}dinger equation w.r.t. $x^{k}$ , $\left(\partial_{k}H_{\text{el}}\right)\ket{u_{0}}=\left(\partial_{k}E_{0}\right)\ket{u_{0}}+(E_{0}-H_{\text{el}})\ket{\partial_{k}u_{0}}$,
and projecting with $Q_{0}$ one easily finds 
\[
Q_{0}\ket{\partial_{k}u_{0}}=G_{0}(E_{0})Q_{0}\left(\partial_{k}H_{\text{el}}\right)\ket{u_{0}}
\]
where $G_{0}(\lambda)=(\lambda-H_{\text{el}}^{0})^{-1}$ is the resolvent
of the \emph{restriction} of $H_{\text{el}}$ to $Q_{0}\mathcal{H}_{\text{el}}$,
\emph{i.e.} the operator $H_{\text{el}}^{0}=Q_{0}H_{\text{el}}=H_{\text{el}}Q_{0}$
defined in the subspace $Q_{0}\mathcal{H}_{\text{el}}$ (here $\mathcal{H}_{\text{el}}$
represents the Hilbert space of the electronic system). More generally,
for $\lambda\in\mathbb{C}$ 
\[
Q_{0}\ket{\partial_{k}u_{0}}=\left(1+(E_{0}-\lambda)G(\lambda)\right)^{-1}G(\lambda)Q_{0}\left(\partial_{k}H_{\text{el}}\right)\ket{u_{0}}
\]
and thus
\begin{align*}
Q_{0}\ket{\partial_{k}u_{0}} & =\left(1+(\lambda-E_{0})G(\lambda)+(\lambda-E_{0})^{2}G(\lambda)^{2}+..\right)\times\\
 & G(\lambda)Q_{0}\left(\partial_{k}H_{\text{el}}\right)\ket{u_{0}}
\end{align*}
provided $\lambda$ is closer to $E_{0}$ than to any other eigenvalue
(here the projector $Q_{0}$ effectively removes the pole at $E_{0}$
in the spectral representation of $G(\lambda)$). If $E_{0}$ is part
of the continuous spectrum we shall use
\[
Q_{0}\ket{\partial_{k}u_{m}}=G^{+}(E_{0})Q_{0}\left(\partial_{k}H_{\text{el}}\right)\ket{u_{0}}
\]
which amounts to defining the eigenvectors $\ket{u_{0}}$ through
a limiting procedure. Specifically, given $H_{\text{el}}$, $E_{0}$
and $\ket{u_{0}}$ at some point $\mathbf{x}$, in order to fix $\ket{u_{0}}$
at a neighboring geometry $\mathbf{x}'=\mathbf{x}+d\mathbf{x}$ one
first defines the family of vectors $\ket{u_{0}^{\lambda}}$ through
the solutions of 
\[
\left[\lambda-(H_{\text{el}}+\Delta H)\right]\ket{u_{0}^{\lambda}}=(\lambda-H_{\text{el}})\ket{u_{0}}
\]
for $\Delta H=\left(\partial_{k}H_{\text{el}}-\partial_{k}E_{0}\right)dx^{i}$.
This gives, for infinitesimal displacements of the nuclear coordinates,
\[
Q_{0}\ket{\partial_{k}u_{0}^{\lambda}}\approx G(\lambda)Q_{0}\left(\partial_{k}H\right)\ket{u_{0}}
\]
from which the above result follows upon taking the limit $\lambda\rightarrow E_{0}$
for $\Im\lambda>0$. 

Consider now the frequency-dependent friction kernel. In the limit
$\omega\rightarrow0$ we can use the above expression for $\lambda=E_{0}+\hbar(\omega+i\epsilon)$
to write 
\begin{align*}
\gamma_{kj}(\omega) & =2i\hbar\lim_{\epsilon\rightarrow0^{+}}\epsilon\int_{0}^{\infty}dte^{-\epsilon t}e^{i\omega t}\times\\
 & \braket{u_{0}|\left(\partial_{k}H_{\text{el}}\right)Q_{0}G(\lambda^{*})e^{-\frac{i}{\hbar}H'_{\text{el}}t}|\partial_{j}u_{0}}\\
\\
 & =-2\hbar^{2}\lim_{\epsilon\rightarrow0^{+}}\epsilon\braket{u_{0}|\left(\partial_{k}H_{\text{el}}\right)Q_{0}G(\lambda^{*})G(\lambda)|\partial_{j}u_{0}}
\end{align*}
and thus, to leading order in $\omega$,
\begin{align*}
\\
\gamma_{kj}(\omega) & =-2\pi\hbar\braket{u_{0}|\left(\partial_{k}H_{\text{el}}\right)\delta(E_{0}+\hbar\omega-H_{\text{el}})|\partial_{j}u_{0}}\\
\end{align*}
where $\lim_{\epsilon\rightarrow0}\hbar\epsilon G(E-i\hbar\epsilon)G(E+i\hbar\epsilon)=\pi\delta(E-H_{\text{el}})$
has been used (here $Q_{0}$ could be removed since the $\delta$
term projects onto states with energy above $E_{0}$). This is precisely
Eq.\ref{eq:Markov friction II} appearing in the main text. 

Equivalently, in the same limit as above, we can also make explicit
the role of $\partial_{j}H_{\text{el}}$ 
\begin{align*}
\gamma_{kj}(\omega) & =-2\pi\hbar\bra{u_{0}}\left(\partial_{k}H_{\text{el}}\right)\delta(E_{0}+\hbar\omega-H_{\text{el}})\\
 & G_{p}(E_{0})\partial_{j}H_{\text{el}}\ket{u_{0}}
\end{align*}
where $G_{p}(E_{0})$ is the principal part of $G^{+}(E_{0})$, that
is, $G_{p}(E_{0})=\lim_{\eta\rightarrow0}\Re G(E_{0}+i\eta)$, and
$\Im G^{+}(E_{0})=-\pi\delta(E_{0}-H_{\text{el}})$ has been neglected
since it gives a vanishing contribution. {[}Here and in the following
$\Re A=(A+A^{\dagger})/2$ and $\Im A=(A-A^{\dagger})/2i${]}. In
the limit $\omega\rightarrow0$ in which we are interested we can
replace the above expression with 
\begin{align*}
\gamma_{kj}(\omega) & =-2\pi\hbar\bra{u_{0}}\left(\partial_{k}H_{\text{el}}\right)\delta(E_{0}+\hbar\omega-H_{\text{el}})\\
 & G_{p}(E_{0}+\hbar\omega)\partial_{j}H_{\text{el}}\ket{u_{0}}
\end{align*}
and thus write the friction kernel as the zero-frequency limit of
\begin{align*}
\gamma_{kj}(\omega) & =\frac{\partial}{\partial\omega}\braket{u_{0}|\left(\partial_{k}H_{\text{el}}\right)\delta(E_{0}+\hbar\omega-H_{\text{el}})\left(\partial_{j}H_{\text{el}}\right)|u_{0}}\\
\end{align*}
where use has been made of 
\begin{align*}
\delta(E-H_{\text{el}})G_{p}(E) & =\frac{1}{2}\lim_{\epsilon\rightarrow0}\epsilon\left[G(\lambda^{*})G(\lambda)^{2}+G(\lambda^{*})^{2}G(\lambda)\right]\\
 & =-\frac{1}{2\hbar}\frac{\partial}{\partial\omega}\lim_{\epsilon\rightarrow0}\epsilon G(\lambda^{*})G(\lambda)\\
 & =-\frac{1}{2\hbar}\frac{\partial}{\partial\omega}\delta(E_{0}+\hbar\omega-H_{\text{el}})
\end{align*}
for $\lambda=E+i\epsilon$, with $E=E_{0}+\hbar\omega$. For comparison
notice that 
\begin{align*}
q_{kj}(\omega) & =-\frac{1}{\hbar}\frac{\partial}{\partial\omega}\braket{u_{0}|\left(\partial_{k}H_{\text{el}}\right)G_{p}(E_{0}+\hbar\omega)\left(\partial_{j}H_{\text{el}}\right)|u_{0}}\\
\end{align*}
gives, in the limit $\omega\rightarrow0$, the quantum geometric tensor. 

Notice further that if the excitation energy $\hbar\omega$ is introduced
at the level of the ``bare'' kernel $\bar{\gamma}_{kj}$, the above
conclusions about the \emph{pseudo}-magnetic contribution and the
residual friction kernel $\gamma_{kj}$ remain unaltered. In fact,
one finds 
\begin{align*}
\bar{\gamma}_{kj}(\omega) & =-2i\hbar q_{kj}+\gamma_{kj}(\omega)\\
 & +2h\omega\braket{\partial_{k}u_{0}|Q_{0}G^{+}(E_{0}+\hbar\omega)|\partial_{j}u_{0}}
\end{align*}
where the last term vanishes in the limit we are interested in.

\subsubsection*{Independent electrons}

For independent electrons $\partial_{k}H_{\text{el}}$ is a monoelectronic
operator that we write as $\partial_{k}h$ and the projector $\delta(E_{0}+\hbar\omega-H_{\text{el}})$
can thus be restricted to singly-excited Slater determinants, \emph{i.e.}
$\ket{\Psi_{a}^{b}}=c_{b}^{\dagger}c_{a}\ket{\Phi_{0}}$ where $\ket{\Phi_{0}}$
is the Hartree-Fock ground-state and $c_{a}$ ($c_{b}^{\dagger}$)
is an annihilation (creation) operator for the single-particle state
$\ket{\phi_{a}}$ ($\ket{\phi_{b}}$). Here, the single-particle energies
are such that $\epsilon_{a}<\epsilon_{F}<\epsilon_{b}$, where $\epsilon_{F}$
is the Fermi level, and $\hbar\omega=\epsilon_{b}-\epsilon_{a}=\Delta\epsilon_{ba}$.
As a result, Eq. \ref{eq:Markov friction II} of the main text becomes
\begin{align*}
\gamma_{kj}(\omega)= & -2\pi\hbar\sum_{a}^{\epsilon_{a}<\epsilon_{F}}\sum_{b}^{\epsilon_{b}>\epsilon_{F}}D_{ab}^{k}\braket{\phi_{b}|\partial_{j}\phi_{a}}\delta(\hbar\omega-\Delta\epsilon_{ba})\\
= & -2\pi\hbar\sum_{a}^{\epsilon_{a}<\epsilon_{F}}\sum_{b}^{\epsilon_{b}>\epsilon_{F}}D_{ab}^{k}D_{ba}^{j}\frac{f(\epsilon_{b})-f(\epsilon_{a})}{\epsilon_{b}-\epsilon_{a}}\delta(\hbar\omega-\Delta\epsilon_{ba})
\end{align*}
where we have defined $D_{ab}^{k}=\braket{\phi_{a}|\partial_{k}h|\phi_{b}}$
and introduced the electron occupation function $f(\epsilon)=\Theta(\epsilon_{F}-\epsilon)$.
Here, we can set $\epsilon_{b}=\epsilon_{a}+\hbar\omega$ in the incremental
ratio of $f$, and taking the limit $\omega\rightarrow0$ replace
it with $-\delta(\epsilon_{a}-\epsilon_{F})$. Hence, upon freeing
the sum over the orbitals we obtain
\[
\gamma_{kj}=\pi\hbar\sum_{a,b}D_{ab}^{k}D_{ba}^{j}\delta(\epsilon_{b}-\epsilon_{F})\delta(\epsilon_{a}-\epsilon_{F})
\]
which is the HGT expression, Eq. \ref{eq:Head-Gordon Tully} of the
main text. Notice that $\gamma_{kj}(\omega)$ above becomes real in
the limit $\omega\rightarrow0$.

\subsubsection*{Frictional vector potential}

Let us show here how the above results follow, in linear response,
by an appropriate modification of the Hamiltonian governing the adiabatic
dynamics, in particular of the vector potential entering such Hamiltonian.
This is important for introducing friction (\emph{i.e.}, dissipation)
into an effective Hamiltonian for the nuclei. As shown below, this
turns the corresponding Schr\"{o}dinger equation into a \emph{non-linear}
equation, but this is the price to pay if the energy transfer mechanism
has to depend on the system dynamics and it is not due simply to an
``external'' field. 

To this end, we work in a \emph{gauge} where the electronic states
evolve according to a zero-averaged Hamiltonian (the ``standard''
\emph{gauge}), \emph{i.e., }$\braket{u^{+}|\partial_{t}u^{+}}=0$
where $^{+}$ denotes the chosen \emph{gauge}. Clearly, in linear-response
this amounts to reference the electronic Hamiltonian to the ground-state
energy, $E_{0}$, and to write $\ket{u^{+}}=\ket{u_{0}}+\ket{\Delta u^{+}}$
where 
\[
\ket{\Delta u^{+}}:=-\frac{i}{\hbar}\int_{0}^{+\infty}e^{-\frac{i}{\hbar}H'_{\text{el}}t'}K_{0}[\psi_{t-t'}]\ket{u_{0}}dt'
\]
since $\braket{u|H|u}\approx E_{0}+2\Re\braket{u_{0}|\Delta u}\equiv E_{0}$
holds thanks to $\braket{\Delta u^{+}|u_{0}}=0$. This also implies
that the nuclear Hamiltonian in this \emph{gauge}
\[
H^{+}\approx\frac{1}{2}\sum_{ij}\xi^{ij}\hat{\pi}_{i}^{+}\hat{\pi}_{j}^{+}+\left(E_{0}(\mathbf{x})+\phi^{+}\right)
\]
resembles closely the \emph{adiabatic} Hamiltonian $H^{0}$: the only
difference is the presence of $\ket{u}=\ket{u_{0}}+\ket{\Delta u^{+}}$
in place of $\ket{u_{0}}$ in the vector and scalar potentials, \emph{e.g.},
\begin{align*}
A_{k}^{0} & =i\braket{u_{0}|\partial_{k}u_{0}}\rightarrow\\
A_{k} & =A_{k}^{0}+i\braket{u_{0}|\partial_{k}\Delta u_{0}}+i\braket{\Delta u_{k}|\partial_{k}u_{0}}\\
 & \equiv A_{k}^{0}+2\Im\braket{\partial_{k}u_{0}|\Delta u_{0}}
\end{align*}
where $\partial_{k}\left(\braket{u_{0}|\Delta u}\right)=0$ has been
used in the last line. In fact, it turns out that the main modification
is precisely the \emph{time-dependent} term
\begin{align*}
\delta A_{k} & =A_{k}-A_{k}^{0}\\
 & =2\Im\left(-\frac{i}{\hbar}\right)\int_{0}^{\infty}\braket{\partial_{k}u_{0}e^{-\frac{i}{\hbar}H'_{\text{el}}t'}K_{0}[\psi_{t-t'}]|u_{0}}dt'
\end{align*}
since this is of first order in the spatial derivative of the electronic
states and generates a force term of the same order through its \emph{time}-derivative,
\begin{align*}
F_{k} & =\frac{\partial\hat{\pi}_{k}^{+}}{\partial t}+\frac{i}{\hbar}[H^{+},\hat{\pi}_{k}]=-\hbar\frac{\partial\left(\delta A_{k}\right)}{\partial t}+\frac{i}{\hbar}[H^{+},\hat{\pi}_{k}^{+}]\\
 & \approx-\hbar\frac{\partial\left(\delta A_{k}\right)}{\partial t}+\frac{i}{\hbar}[H^{0},\hat{\pi}_{k}^{0}]
\end{align*}
where $\hat{\pi}_{k}^{0}=\hat{p}_{k}-\hbar A_{k}^{0}$ is the adiabatic
momentum. Here, the last line holds if we retain only terms that contain
up to three spatial derivatives of the electronic state at a time,
\emph{e.g.}, of the form 
\[
-\partial_{k}\phi_{0}=-\frac{\hbar^{2}}{2}\sum_{ij}\xi^{ij}\partial_{k}\Re\braket{\partial_{i}u_{0}|Q_{0}|\partial_{j}u_{0}}
\]
In other words, to this ``order'' in the spatial derivatives, we
have
\[
[\hat{\pi}_{i}^{+},\hat{\pi}_{j}^{+}]\approx[\hat{\pi}_{i}^{0},\hat{\pi}_{j}^{0}]\ \ \phi^{+}\approx\phi^{0}
\]
which can be summarized by stating that the geometric properties are
the same as in the adiabatic limit, $q_{ij}^{+}\approx q_{ij}^{0}$.

Let us then take a closer look at the correction $\delta A_{k}$ to
the vector potential. From the definition of $K_{0}$ we have 
\begin{align*}
\delta A_{k} & =-2\sum_{j}\Im\int_{0}^{\infty}\braket{\partial_{k}u_{0}|e^{-\frac{i}{\hbar}H'_{\text{el}}t'}Q_{0}|\partial_{j}u_{0}}V_{t-t'}^{j}dt'\\
 & +2\Re\int_{0}^{\infty}\braket{\partial_{k}u_{0}|e^{-\frac{i}{\hbar}H'_{\text{el}}t'}R|u_{0}}dt'
\end{align*}
where the second term can be neglected in the approximation above
since it is time independent and it is of third order, hence contributes
to the force only with a fourth order term. Upon introducing the complex-valued
``position'' fields
\[
X^{j}(\mathbf{x},t)=\int_{-\infty}^{t}V^{j}(\mathbf{x},t')dt'
\]
and integrating by parts we find 
\begin{align*}
\delta A_{k} & =-2\sum_{j}\Im\left(q_{kj}^{0}X^{j}\right)\\
 & +\frac{2}{\hbar}\sum_{j}\Re\left(\int_{0}^{\infty}\braket{\partial_{k}u_{0}|\Gamma(t')|\partial_{j}u_{0}}X_{t-t'}^{j}dt'\right)
\end{align*}
where $\Gamma(t)=H_{\text{el}}'e^{-\frac{i}{\hbar}H'_{\text{el}}t}Q_{0}$
has been introduced. In the Markov limit we have
\begin{align*}
\int_{0}^{\infty}\braket{\partial_{k}u_{0}|\Gamma(t')|\partial_{j}u_{0}}X_{t-t'}^{j}dt & \approx\\
X_{t}^{j} & \int_{0}^{\infty}\braket{\partial_{k}u_{0}|\Gamma(t')|\partial_{j}u_{0}}dt
\end{align*}
where, as seen above, 
\[
\lim_{\epsilon\rightarrow0^{+}}\int_{0}^{\infty}e^{-\epsilon t}\Gamma(t)dt=-i\hbar\left(1+i\pi H_{\text{el}}^{'}\delta(H_{\text{el}}^{'})\right)Q_{0}
\]
Hence, in this limit, we find the following simple ``frictional correction''
to the adiabatic dynamics
\[
\delta A_{k}=2\pi\sum_{j}\Re\left(\braket{\partial_{k}u_{0}|H'_{\text{el}}\delta(H'_{\text{el}})|\partial_{j}u_{0}}X_{t}^{j}\right)
\]
It is not hard to show that these expressions for $\delta A_{k}$
give exactly the correction to the forces discussed above. In the
Markov limit, for instance, $F_{k}=F_{k}^{0}+\delta F_{k}$ where
\[
\delta F_{k}=-\sum_{j}\gamma'_{kj}\Re V_{t}^{j}+\sum_{j}\gamma''_{kj}\Im V_{t}^{t}
\]
and $\gamma'$ and $\gamma''$ are, respectively, the real and imaginary
parts of the kernel 
\[
\gamma_{kj}=2\pi\hbar\braket{\partial_{k}u_{0}|(H'_{\text{el}}\delta(H'{}_{\text{el}})|\partial_{j}u_{0}}
\]

In the Markov limit, in the simplest case where $\xi^{ij}=\delta_{ij}M^{-1}$,
if $\gamma_{kj}$ can be taken \emph{diagonal} and \emph{uniform}
in the configuration space of the system where the dynamics occurs,
we can write
\[
\delta A_{k}=\Re(\gamma X_{t}^{k})\approx\partial_{k}\Re\left(-i\hbar M^{-1}\gamma\int_{-\infty}^{t}\ln\psi_{t'}(\mathbf{x})dt'\right)
\]
if we neglect the contribution of the vector potential to the velocity
field. We thus see that $\delta A_{k}$ becomes longitudinal and can
be replaced by an appropriate \emph{scalar} field
\[
\delta\phi=\hbar M^{-1}\Im\left(\gamma\ln\psi_{t}(\mathbf{x})\right)
\]
When setting $\gamma''\approx0$ the resulting Hamiltonian $H=H^{0}+\delta\phi$
is the Kostin\textbf{ }Hamiltonian which is a simple way to introduce
dissipation into a Schr\"{o}dinger-like equation by adding a simple
``phase potential'' (depending on the phase of the system wavefunction
in the position representation). 
\end{document}